\documentclass{article}
\usepackage{cite}
\usepackage{amsmath}
\begin{document}
\title{On the uncertainty relations and quantum measurements: conventionalities, 
shortcomings, reconsiderations}
\author{S. Dumitru \footnote{Department of Physics, "Transilvania" University, B-dul Eroilor 29, Brasov,  Romania}}
\maketitle

\begin{abstract}
Unsolved controversies about uncertainty relations and quantum measurements still persists 
nowadays. They originate around the shortcomings regarding the conventional 
interpretation of uncertainty relations. Here we show that the respective shortcomings 
disclose veridic and unavoidable facts which require the abandonment of the mentioned 
interpretation. So the primitive uncertainty relations appear as being either thought 
fictions or fluctuations formulae. Subsequently  we reveal that the conventional approaches
of quantum measurements are grounded on incorrect premises. We propose a new approach in 
which : (i) the quantum observables are considered as generalized stochastic variables, 
(ii) the view is focused only on the pre-existent state of the measured system, without 
any interest for the collapse of the respective state, (iii) a measurement is described 
as an input-output transformation which modify the probability density and current but 
preserve the expressions of the operators. The measuring uncertainties are evaluated as 
changes in the probabilistic estimators of observables. Related to different observables 
we do not find reasons of principle neither for uncertainties-connections  nor for  
measuring compatibility/incompatibility.


\end{abstract}

\section{Introduction}\label{intro}
Debates about the uncertainty relations (UR) are present in a large number of early as 
well as recent publications (for a significant bibliography see \cite{1,2,3,4,5,6,7}).
In a direct or dissimulated manner most of the respective debates are connected with 
the so-called conventional interpretation of UR (CIUR), promoted by the Copenhagen School and its partisans. 
Often CIUR is mentioned as fragmentary  excerpts from a diffuse collection of statements. 
For all that it can be  confined  around a restricted number of basic ideas (see below the next section). 
Less mentioned is the fact that CIUR ideas are troubled by a number of still unsolved 
shortcomings. As a rule, in the main stream of recent publications, the alluded 
shortcomings are underestimated (through unnatural solutions or even by omission).

The most known shortcomings of CIUR regards \cite {1,8,9,10,11,12,13,14,15,16,17,18,19,20,21,22,23,24} 
the following pairs of canonically conjugated observables $L_z  - \varphi$, $N - \phi$ and $E - t$  
($L_z$ = $z$ component of angular momentum, $\varphi$ = azimuthal angle, $N$ = number, 
$\phi$ = phase, $E$ = energy, $t$ = time). The respective pairs show anomalies in respect
with the usual (Robertson - Schrodinger) version of theoretical UR and, consequently, 
they are in conflict with CIUR ideas. For solving the conflict in literature the current
attitude is to preserve CIUR as an immovable doctrine and to adjust adequately the 
expressions of theoretical UR (for details see  the Sections \ref{sec3} and \ref{sec4}). But as an
intriguing fact the mentioned adjustments differ among them both quantitatively and 
qualitatively. So, until now, an agreement regarding the mentioned conflict does not exist 
in the scientific literature.

In the alluded circumstances we think that an investigation ab origine of the facts 
is of major interest. Such an investigation has to include: 

(i) firstly, a search of the primary mathematical source regarding the theoretical UR and \\
(ii) secondly, a consideration of the respective source as a standard (reference element) in the appreciation of 
things. 

An investigation of the mentioned kind we propose  in Section \ref{sec5}. Thus
we find that the searched source/standard is a Cauchy-Schwarz relation that must replace the 
usual UR in debates regarding the above mentioned  pairs of observables. But the 
respective  finding disclose insurmountable contradictions with CIUR.

A good appraisal of the proposed investigation requires to examine also other things 
which reveal additional shortcomings of CIUR. Some of the respective things were mentioned 
in publications only occasionally and without noticeable impact in scientific community.
But a significant class of such things can be collected by a careful exploration of the 
corresponding facts and literature (see  Section \ref{sec6}). A minute examination of the 
elements of the respective class shows that each of them contradict in an indubitable 
manner one or more basic ideas (\textbf{\emph{BI}}) on which CIUR  relies. Moreover the ensemble 
of the respective contradictions, together with the inherent anomalies shown by the above precised pairs of 
conjugated observables, incriminate in an insurmountable way the whole set of the 
mentioned BI. So CIUR appears in a situation of indubitable failure and, consequently,
it must be re-evaluated and abandoned as an incorrect doctrine (for details 
see Section \ref{sec7}).

The mentioned re-evaluation of CIUR entails a complete disconnection of UR from their 
supposed significance regarding the quantum measurements (QMS). The description of QMS is 
a question approached in a large number of controversial debates (see publications 
\cite{1,2,3,4,5,6,25,26,27,28,29} and references). Most of the respective debates 
promoted the  presumption that the alluded description must be included within the 
framework of quantum mechanics (QM). Now we note that, in spite of the mentioned 
disconnection of UR, the description of QMS must remain a natural subject for scientific
investigations. On the other hand we think that the above alluded old presumption is 
unjustified from the viewpoint of real physics practice . We opine that, according to the  
respective practice, the description of QMS must be considered as a distinct and
additional task in respect with the purposes of usual QM. Our opinions regarding the 
reconsideration of QMS description are argued, presented and exemplified in Sections \ref{sec8}-\ref{sec11} 
and in Annex A.

\section{Basic ideas of CIUR} \label{sec2}
In the main CIUR concerns on the purpose to give a unique and generic interpretation for
the thought-experimental (\emph{te}) relation
\begin{equation}\label{eq:1}
\Delta _{te} A \cdot \Delta _{te} B \ge \hbar 
\end{equation}
and for the theoretical formula
\begin{equation}\label{eq:2}
\Delta _\psi  A \cdot \Delta _\psi  B \ge \frac{1}{2}\left| {\left\langle 
{\left[ {\hat A,\hat B} \right]} \right\rangle _\psi  } \right|
\end{equation}
Relation \eqref{eq:1} regards the \emph{te}-uncertainties $\Delta _{te} A$ and
$\Delta _{te} B$  for conjugated observables $A$ and $B$ (like $A = x$ = coordinate  and
$B = p$ = momentum ). It was introduced \cite{30,31} by means of so-called 
thought (or mental) experiments. Formula \eqref{eq:2} (known as Robertson - Schr\"odinger UR) derives
from the mathematical formalism of QM (for some details see also  Section \ref{sec5}).

Motivated by the mentioned concern CIUR was widely popularized and, explicitly or 
implicitly, it is agreed in a large number of old as well as recent publications. But as a 
strange aspect, in dissonance with such facts, in its partisan literature CIUR is often 
presented so vaguely and fragmentarily that it seems to be rather an intricate conception 
but not a well defined doctrine. However, in spite of the respective aspect, one can see
\cite{32,33,34} that in fact CIUR is founded on a resticted number of basic 
ideas (\textbf{\emph{BI}}).   Here we report the respective \textbf{\emph{BI}} as follows:

\emph{\textbf{\underline{BI.1}}: Quantities $\Delta _{te} A$ and $\Delta _\psi  A$ from
relations \eqref{eq:1} and \eqref{eq:2}, denoted by a unique symbol $\Delta A$, 
have similar significance of measuring uncertainty for the observable$A$. Consequently 
the respective relations have the same generic interpretation as UR regarding the 
simultaneous measurements of observables $A$ and $B$.} 

\emph{\textbf{\underline{BI.2}}: In case of a solitary observable $A$ the quantity
$\Delta A$  always can have an unbounded small value. Therefore such an obvservable 
can be measured without uncertainty in all cases of systems and states}.

\emph{\textbf{\underline{BI.3}}: When two observables $A$ and $B$ are commutable (i.e
$\left[ {\hat A,\,\hat B} \right]\;\, = \;\,0$) relation \eqref{eq:2} allows for 
the quantities $\Delta A$ and $\Delta B$ to be unlimitedly small at the same time. That is why such observables
can be measured simultaneously and without uncertainties for any system  or state. 
Therefore they are considered as compatible}.

\emph{\textbf{\underline{BI.4}}: If two observables $A$ and $B$ are noncommutable 
(i.e. $\left[ {\hat A,\,\hat B} \right]\neq 0$) relation \eqref{eq:2} shows that the 
quantities 
 $\Delta A$ and $\Delta B$ can be never reduced concomitantly to null values. For that 
 reason such observables can be measured simultaneously only with non-null and 
 interconnected uncertainties, irrespective of the system or state.Consequently such
 observables are considered as incompatible}.

 \emph{\textbf{\underline{BI.5}}: Relations \eqref{eq:1} and \eqref{eq:2}, Planck's
 constant $\hbar$ as well as the measuring peculiarities noted in \textbf{BI.4 }
 are typically QM things which have not analogy in classical (non-quantum) macroscopic
 physics }.

\section{Cases of angular observables}\label{sec3}
In its integrity (previously delimited through \emph{\textbf{BI.1-5}}) CIUR is vulnerable 
to shortcomings connected with the pairs of angular observables $L_z - \varphi$ and 
$N - \phi$ (z-component of angular momentum - azimuthal angle, respectively 
number - phase). Some of the respective shortcomings were debated in publications from
the last decades (see \cite{1,8,9,10,11,12,13,14,15,16,17,18} and references).

As regards the $L_z - \varphi$ pair the mentioned debates revealed the following facts:
According to the usual procedures of QM, $L_z$ and $\varphi$ should be described by the 
conjugated operators
\begin{equation}\label{eq:3}
\hat L_z  =  - i\hbar \frac{\partial }{{\partial \varphi }} ,\quad \quad 
\hat \varphi  = \varphi  \cdot 
\end{equation}
respectively by the commutation relation
\begin{equation}\label{eq:4}
\left[ {\hat L_z , \hat \varphi } \right] = - i\hbar 
\end{equation}
So for the alluded pair the CIUR's basic formula \eqref{eq:2} requires directly 
the relation
\begin{equation}\label{eq:5}
\Delta_\psi  L_z \cdot \Delta_\psi \, \varphi \ge \frac{\hbar}{2}
\end{equation}
On the other hand the angular states of some systems (which will be specified below) are
described by the wave functions
\begin{equation}\label{eq:6}
\psi _m (\varphi ) = (2\pi )^{ - \frac{1}{2}} \, e^{im\varphi} 
\end{equation}
with $\varphi \in [0, 2 \pi )$ and $m = 0, \pm 1, \pm 2,\ldots$. 
For such states one obtains
\begin{equation}\label{eq:7}
\Delta _\psi  L_z  = 0, \quad \quad \Delta _\psi \,\varphi = \frac{\pi }{{\sqrt 3 }}
\end{equation}
But these expressions are incompatible with  relation \eqref{eq:5}. 

For avoiding the mentioned incompatibility many publications promoted the conception 
that in the case of $L_z - \varphi$  pair the usual procedures of QM do not work correctly.
Consequently it was accredited the idea that formula \eqref{eq:5} must be prohibited 
and replaced by adjusted $\Delta _\psi L _z - \Delta _\psi\, \varphi$ relations resembling
with \eqref{eq:2}. So, along the years, a lot of such adjusted relations were proposed.
In the main the respective relations are expressed in one of the following forms:
\begin{equation}\label{eq:8}
\Delta _\psi  L_z  \cdot \Delta _\psi  f(\varphi ) \ge \hbar \,
\left| {\left\langle {g(\varphi )} \right\rangle _\psi  } \right|
\end{equation}  
\begin{equation}\label{eq:9}
\left( {\Delta _\psi  L_z } \right)^2  + \hbar ^2 \left( {\Delta _\psi u(\varphi )} \right)^2 \ge \hbar ^2 
\left\langle {v(\varphi )} \right\rangle _\psi ^2 
\end{equation} 
\begin{equation}\label{eq:10}
\Delta _\psi  L_z \cdot \Delta _\psi  \,\varphi \ge 
\frac{\hbar }{2} \left| {1 - 2\pi \left| {\psi (2 \pi  - 0)} \right|} \right|
\end{equation} 
where $\psi (2 \pi - 0) := \lim \limits_{\varphi \to 2 \pi  - 0} \psi(\varphi)$.

In \eqref{eq:8} - \eqref{eq:9} $f(\varphi)$,  $g(\varphi)$, $u(\varphi)$ and $v(\varphi)$ denote various
adjusting functions of $\varphi$ asserted by means of some circumstantial (and more or
less fictitious) considerations.

A minute examination of the facts shows that, in essence, the relations \eqref{eq:8} -
\eqref{eq:10} are troubled by shortcomings revealed within the following 
remarks (\textbf{R}).

\textbf{\underline{R.1}}: None of the respective relations is agreed unanimously as a
correct $\Delta _\psi L_z - \Delta _\psi \varphi$ relation able to replace  formula
\eqref{eq:5}. 

\textbf{\underline{R.2}}: Mathematically the mentioned relations are not mutually 
equivalent.

\textbf{\underline{R.3}}: Relations \eqref{eq:8} - \eqref{eq:9} have no rational 
justifications in the usual formalism of QM (that however works very well in a huge
number of applications).

\textbf{\underline{R.4}}: The relation \eqref{eq:10} is correct from the usual QM 
perspective (see  formula \eqref{eq:41} in Section 5) but it conflicts with the 
CIUR's idea  \textbf{\emph{BI.4}} in the case of states described by the wave functions
\eqref{eq:6}.

In respect with the $N-\phi$ pair  the situation is as follows. The respective pair 
refers to a quantum oscillator and it is described by the operators  
$\hat N $  and $\hat \phi$ introduced by relations
\begin{equation}\label{eq:11}
\hat a = e^{i\hat \phi } \sqrt {\hat N} \,, \quad \quad \hat a^ +   = \sqrt {\hat N} 
e^{ - i\hat \phi } 
\end{equation}
where $\hat a $ and ${\hat a^+}$ denote 
the known ladder (lowering and raising) operators. From \eqref{eq:11} one finds
\begin{equation}\label{eq:12}
\left[ {\hat N, \hat \phi } \right] = i
\end{equation}
CIUR's basic formula \eqref{eq:2} requires thus directly the relation
\begin{equation}\label{eq:13}
\Delta _\psi  N \cdot \Delta _\psi \phi  \ge \frac{1}{2}
\end{equation}
On the other hand for an oscillator in an energetic eigenstate one obtains the results
\begin{equation}\label{eq:14}
\Delta _\psi  N = 0 \, ,\quad \quad \Delta _\psi \phi  = \frac{\pi }{{\sqrt 3 }}
\end{equation}
(the last of these results, not mentioned explicitly in many publications, can be 
obtained easily by means of the wave functions \eqref{eq:17} given below). Now one can 
see that the results \eqref{eq:14} invalidate the relation \eqref{eq:13} and so it is 
revealed the anomaly of $N-\phi$ pair in respect with CIUR.

Here we  also add the next remarks.

\textbf{\underline{R.5}}: Mathematically the situation of the $N-\phi$ pair is completely
similar with that of the $L_z-\varphi$ pair. The respective similarity can be pointed 
out as follows.  If the wave functions are considered in the $\phi$-representation (with
$\phi \in [0, 2\pi)$) from \eqref{eq:12} results that the operators  $\hat N$ and 
$\hat \phi$ have the expressions
\begin{equation}\label{eq:15}
\hat N = i\frac{\partial }{{\partial \phi }}\,,\quad \quad \hat \phi = \phi  \cdot 
\end{equation}
Then the Schr\"odinger equation for oscillator take the form
\begin{equation}\label{eq:16}
\hbar \omega \left( {i\frac{\partial }{{\partial \phi }} + \frac{1}{2}} \right)\psi = E\psi 
\end{equation}
where $E$ = energy and $\omega$ = angular frequency.

By considering $\psi (2 \pi-0) = \psi(0)$ and $E>0$ it results that in the 
$\phi$-representation a quantum oscillator is described by the wave functions
\begin{equation}\label{eq:17}
\psi _N \left( \phi  \right) = \left( {2\,\pi } \right)^{ - \frac{1}{2}} e^{ - iN\phi } 
\end{equation}
with $N = 0, 1,2, 3,\ldots$  (correspondingly $E = E_N = \hbar\omega (N + 1/2)$). 
Then a direct comparison of pairs of relations \eqref{eq:15} and \eqref{eq:17} 
respectively \eqref{eq:3} and \eqref{eq:6} attests the announced mathematical similarity between 
the pairs $N-\phi$ and $L_z -\varphi$.

\textbf{\underline{R.6}}: The similarity between the two pairs is also evidenced (perhaps
less visible) by the various $\Delta _\psi N - \Delta _\psi \phi$ adjusted formulae 
proposed in literature (see \cite{11,13,14,16}) in order to replace \eqref{eq:13} and to
avoid the $N - \phi$ anomaly in respect with CIUR. In their essence the respective
formulae are completely analogous with the relations \eqref{eq:8} - \eqref{eq:10} for the
$L_z - \varphi$ pair. Moreover, it is easy to see that the mentioned formulae are troubled 
by shortcomings which are similar with the ones mentioned above in \textbf{R.1 - 4}.

Now let us note that in the $L_z - \varphi$ case the states described by \eqref{eq:6} regards
exclusively the restricted class of sharp circular rotations (SRC) around the z-axis. Such
SRC are specific for a particle on a circle, for a 1D (or fixed-axis) rotator, and 
for non-degenerate spherical rotations respectively. One finds examples of systems with 
spherical rotations in the cases of a particle on a sphere, of 2D or 3D rotators and  
of an electron in a hydrogen atom respectively. The mentioned rotations are considered 
as non-degenerate if all the specific (orbital) quantum numbers have well-defined (unique) values.

Here is the place to remind  that the situation of the $L_z - \varphi$ pair must be also
discussed  in relation with the extended rotations (EXR). By EXR we refer to the 
quantum torsion pendulum (QTP) and  to the degenerate spherical rotations (of the
above mentioned systems) respectively. A rotation (motion) is degenerate if the energy 
of the system is well-specified while the non-energetic quantum numbers (here of orbital nature) take
all  permitted values.

From the class of EXR let us firstly refer to the case of a QTP which \cite{33,34}
can be regarded as a simple QM oscillator. Indeed a QTP which oscillate around the z-axis 
is characterized by the Hamiltonian 
\begin{equation}\label{eq:18}
\hat H = \frac{1}{{2I}} \hat L_z^2  + \frac{1}{2} I\omega ^2 \varphi ^2 
\end{equation}
Here $\varphi$ denotes the azimuthal angle with $\varphi \in (-\infty, \infty)$, $\hat{L}_z$
is the $z$-component of angular momentum operator defined by \eqref{eq:3}, $I$ is 
the momentum of inertia and $\omega$ represents the (angular) frequency of torsion oscillation. 
Then the eigenstates of QTP have   energies $E_N = \hbar\omega(N + 1/2)$ and are described 
by the wave functions
\begin{equation}\label{eq:19}
\psi _N (\varphi ) = \psi _N (\xi ) \propto \exp \left( { - \frac{{\xi ^2 }}{2}} \right)
\mathcal{H}_N (\xi )\,,\quad \xi  = \varphi \sqrt{\frac{{I\omega }}{\hbar }} 
\end{equation}
where $N = 0, 1, 2, 3,\ldots$ signifies the oscillation quantum number and $\mathcal{H}_N(\xi)$ 
stand for Hermite polinomials of $\xi$. In the states \eqref{eq:19} for the observables $L_z$
and $\varphi$ associated with the operators \eqref{eq:3} one obtains
\begin{equation}\label{eq:20}
\Delta _\psi  L_z  = \sqrt {\hbar I\omega \left( {N + \frac{1}{2}} 
\right)} \, , \quad \quad \Delta _\psi  \varphi  = \sqrt {\frac{\hbar }
{{I\omega }}\left( {N + \frac{1}{2}} \right)} 
\end{equation}
With these expressions for $\Delta _\psi L_z$ and $\Delta _\psi \varphi$ one finds 
that for the considered QTP the $L_z - \varphi$ pair satisfies the prohibited formula
\eqref{eq:5}.

From the same class of EXR let us now refer to a degenerate state of a particle on a 
sphere or of a 2D rotator. In such a state the energy is $E= \hbar^2 l (l+ 1)/2I$ 
where the orbital number $l$ has a well-defined value ($I$ = moment of 
inertia). In the same state the magnetic number $m$ can take all the values $-l, -l+1,
\ldots, -1, 0, 1, \ldots,l-1, l$. 
Then the mentioned state is described by a wave function of the form
\begin{equation}\label{eq:21} 
\psi _l (\theta , \varphi ) = \sum\limits_{m =  - l}^l {c_m } \,Y_{lm} (\theta ,\varphi )
\end{equation}
Here $\theta$ and $\varphi$ denote the polar and  azimuthal angle respectively 
( $\theta \in  [0,\pi],  \varphi \in [0, 2\pi)$), $Y_{lm} \,(\theta ,\varphi )$ are 
the spherical functions and $c_m $ represent complex coefficients which satisfy the normalization 
condition $\sum\limits_{m =  - l}^l {\left| {c_m } \right|^2 } = 1$. With the
expressions \eqref{eq:3} for the operators $\hat L _z $ and $\hat \varphi $ in a state described 
by \eqref{eq:21} one obtains
\begin{equation}\label{eq:22}
\left( \Delta_\psi   L_z  \right)^2 = \sum\limits_{m =  - l}^l {\left| {c_m } \right|^2 } \,
\hbar^2 \,  m^2  - \left[ \sum_{m =  - l}^l \left| {c_m } \right|^2 \hbar m  \right]^2 
\end{equation}
\begin{equation}\label{eq:23}
\begin{split}
\left( \Delta_\psi  \,\varphi  \right)^2  &= \sum\limits_{m =  - l}^l 
\sum\limits_{k =  - l}^l c_m^*  \, c_k \left( Y_{lm} , \varphi^2 \, Y_{lk}  \right) -  \\
&- \left[ \sum\limits_{m =  - l}^l \sum\limits_{r =  - l}^l c_m^* \, c_r  \left(
Y_{lm} ,\varphi Y_{lr}  \right)  \right]^2   
\end{split}
\end{equation}
where $(f, g)$ denotes the scalar product of the functions $f$ and $g$.

By means of the expressions \eqref{eq:22} and \eqref{eq:23} one finds that in the case of 
alluded degenerated EXR, described by \eqref{eq:21} it is possible for the prohibited
formula \eqref{eq:5} to be satisfied. Such a possibility is conditioned by the concrete 
values of the  $c _m$ coefficients.

\textbf{\underline{R.7}}: The facts presented above in this section prove 
that, in reality, CIUR is unable to give a natural and unitary approach  of the problems 
connected with the pairs of angular observables $L _z - \varphi$ and $N - \phi$. The
respective unableness can not be remedied in a way by means of the inner resources of CIUR.
Consequently it must be recorded as a major and unavoidable shortcoming of CIUR.

\section{The case of energy and time} \label{sec4}
Another pair of (canonically) conjugated observables which are unconformable in relation 
with the CIUR ideas is given by energy $E$ and time $t$. That is why the respective pair 
was the subject of a large number of (old as well as recent) controversial discussions
(see \cite{20,21,22,23,24} and references). The alluded discussions were generated by 
the following observations. On  one hand $E$ and $t$, as  conjugated observables  have 
to be described in  terms of QM  by the operators
\begin{equation}\label{eq:24}
\hat E = i\hbar \frac{\partial }{{\partial t}}
\, , \quad \quad \hat t= t \cdot 
\end{equation}
respectively by the commutation relation
\begin{equation}\label{eq:25}
\left[ \hat E , \hat t \right] = i\hbar 
\end{equation}
In accordance with \eqref{eq:2} such description require the relation
\begin{equation}\label{eq:26}
\Delta _\psi  E \cdot \Delta _\psi t  \ge \frac{\hbar }{2}
\end{equation}
On the other hand because in usual QM the time $t$ is a deterministic   but not a 
stochastic variable for any quantum situation (system and state) one finds the expressions
\begin{equation}\label{eq:27}
\Delta _\psi  E = \text{a finite quantity}\, , \quad \quad \Delta _\psi t \equiv 0
\end{equation}
But these expressions invalidate the relation \eqref{eq:26} and consequently show an
anomaly in respect with the CIUR ideas (especially with \textbf{\emph{BI.4}}). For avoiding
the alluded anomaly CIUR partisans invented a lot of adjusted $\Delta E - \Delta t$ 
formulae destined to substitute the questionable relation \eqref{eq:26}(see \cite{1,20,21,22,23,24} and
references). The mentioned formulae can be written in the generic form
\begin{equation}\label{eq:28}
\Delta _v E \cdot \Delta _v t \ge \frac{\hbar}{2}
\end{equation}
Here $\Delta _v E$ and $\Delta _v t$ have various ($v$) significances such as: 
(i) $\Delta _1 E$ = line-breadth of the spectrum characterizing  the  decay of an excited 
state  and $\Delta _1 t$ = half-life of the respective state, (ii) $\Delta _2 E$ = 
$\hbar \Delta \omega$ = spectral width (in terms of frequency $\omega$)
of a wave packet and  $\Delta _2 t$ = temporal width of the wave packet, 
(iii) $\Delta _3 E = \Delta _\psi E$ and  $\Delta _3 t = \Delta _\psi A 
\cdot \left( {d\left\langle A \right\rangle _\psi  /dt} \right)^{ - 1}  $, with $A$ = an 
arbitrary observable.

Note that in spite of the efforts and imagination implied in the disputes connected 
with the formulae \eqref{eq:28} the following remarks remain of topical interest.

\textbf{\underline{R.8}}: The diverse formulae from the family \eqref{eq:28} are not 
mutually equivalent from a mathematical viewpoint. Moreover they have no natural 
justification in the framework of usual QM (that however give a huge number of good 
results in applications).

\textbf{\underline{R.9}}: In the specific literature (see \cite{1,20,21,22,23,24} 
and references) none of the formulae \eqref{eq:28} is agreed unanimously as a correct 
substitute for  relation \eqref{eq:26}.

\textbf{\underline{R.10}}: Consequently the applicability of the CIUR ideas to the $E - t$
pair persists in our days as a still unsolved question.

\section{An investigation  ab origine of the facts }\label{sec5}
In its essence the above presented conflict of CIUR with the  mentioned pairs of 
observables regards the applicability of the theoretical formula \eqref{eq:2}. A
clearing up of the facts requires an investigation ab origine of the  conditions/range
of validity for the respective formula. Such an investigation can be done as follows.

Let us consider a quantum system whose state and observables $A _j$ ($j = 1, 2, \ldots, r$) 
are described by the wave function $\psi$ and by the operators 
$\hat A_j  $ respectively. If $(f,g)$ denote the scalar product of the functions
$f$ and $g$ the quantity $\left\langle A_j  \right\rangle = \left( {\psi \,, \hat A_j \psi } \right)$ 
represents the mean (expected) value of the observable $A_j$ in the mentioned state. 
Because $A_j$ are stochastic variables they show 
fluctuations (deviations from the mean values).  The respective fluctuations are 
described (in a first order  approximation) by means of the correlations
$C_{jk} = \left( {\delta _\psi  \hat A_j \psi ,\, \delta _\psi  \hat A_k \psi } 
\right)$ where $\delta _\psi  \hat A_j = \hat A_j  - \left\langle {A_j }\right\rangle $. 
It is easily to see that the alluded correlations satisfy the following two relations
\begin{equation}\label{eq:29}
\left( {\delta _\psi  \hat A_j \psi ,\, \delta _\psi  \hat A_k \psi } \right)^* =
\left( {\delta _\psi  \hat A_k \psi ,\, \delta _\psi  \hat A_j \psi } \right)
\end{equation}
\begin{equation}\label{eq:30}
\left\| {\sum\limits_{j = 1}^r {\lambda _j \delta _\psi  \hat A_j \psi } } 
\right\|^2 = \sum\limits_{j = 1}^r {\sum\limits_{k = 1}^r {\lambda _j^* } } 
\lambda _k \left( {\delta _\psi  \hat A_j \psi ,\,\delta _\psi  \hat A_k \psi } 
\right) \ge 0
\end{equation}
which imply the notations: $f^* $ =  complex conjugate of $f$, $\left\| g \right\|$ = 
norm of $g$, $\lambda _j (j = 1, 2, \ldots, r)$ = a set of arbitrary and complex parameters. 
The relations \eqref{eq:29} and \eqref{eq:30} show that the set of correlations $C_{jk}$
constitutes a Hermitian and non-negatively defined matrix. Then in accordance with the 
matrix algebra \cite{36} can be written the formula 
\begin{equation}\label{eq:31}
\det \left[ {C_{jk} } \right]= \det \left[ {\left( {\delta _\psi  \hat A_j 
\psi ,\,\delta _\psi  \hat A_k \psi } \right)} \right] \ge 0
\end{equation}
where $\det \left[ {C_{jk} } \right]$ denotes the determinant with elements $C _{jk}$. For
two observables $A _1 = A$ and $A _2 = B$ from \eqref{eq:31} one obtains
\begin{equation}\label{eq:32}
\left( {\delta _\psi  \hat A \psi ,\,\delta _\psi  \hat A\psi } \right)\left( 
{\delta _\psi  \hat B\psi ,\,\delta _\psi  \hat B\psi } \right) \ge \left| 
{\left( {\delta _\psi  \hat A\psi ,\,\delta _\psi  B\psi } \right)} \right|^2 
\end{equation}
i.e. a Cauchy-Schwarz relation for the functions $\delta _\psi \hat A \psi$ and  
$\delta _\psi \hat B \psi$.
For an observable $A$ regarded as a stochastic variable the quantity $\left( {\delta _\psi  
\hat A \psi ,\,\delta _\psi  \hat A \psi } \right)^{\frac{1}{2}} = \Delta _\psi A$ 
represents its standard deviation. From \eqref{eq:32} it results directly that the
standard deviations $\Delta _\psi A$ and  $\Delta _\psi B$ of two observables $A$ and $B$
satisfy the relation
\begin{equation}\label{eq:33}
\Delta _\psi  A \cdot \Delta _\psi  B \ge \left| {\left( {\delta _\psi  
\hat A \psi ,\,\delta _\psi  B \psi } \right)} \right|
\end{equation}
which can be called \emph{Cauchy-Schwarz formula}. Note that the relations \eqref{eq:31} - 
\eqref{eq:33} are always valid  (i.e. for all observables, systems and states). The 
formula \eqref{eq:33} implies the less general  UR \eqref{eq:2} only when the two 
operators $\hat A  = \hat A_1$ and $\hat B = \hat A_2$ satisfy the conditions
\begin{equation}\label{eq:34}
\left( {\hat A_j \psi ,\hat A_k \psi } \right) = \left( {\psi ,\hat A_j 
\hat A_k \psi } \right)\quad \quad \left( {j = 1,2;\;k = 1,2} \right)
\end{equation}
Indeed in such cases one can write the relation
\begin{equation}\label{eq:35}
\begin{split}
\left( {\delta _\psi  \hat A\psi , \, \delta _\psi  \hat B\psi } \right) 
=& \dfrac{1}{2} \left( {\psi , \, \left( {\delta _\psi  \hat A \cdot \delta _\psi 
\hat B\psi  + \delta _\psi  \hat B \cdot \delta _\psi  \hat A} \right)\psi } \right) -  \\  
&- \dfrac{i}{2}\left( {\psi , \, i\left[{\hat A, \, \hat B} \right]} \right) 
\end{split}
\end{equation}
where the two terms from the right hand side are purely real and  imaginary  
quantities respectively. Therefore in the mentioned cases from \eqref{eq:33} one finds
\begin{equation}\label{eq:36}
\Delta _\psi  A \cdot \Delta _\psi  B \ge \frac{1}{2}\left| 
{\left\langle {\left[ {\hat A,\, \hat B} \right]} \right\rangle _\psi  } \right|
\end{equation}
i.e. the usual UR \eqref{eq:2}. The above general framing of UR \eqref{eq:2}/\eqref{eq:36} 
suggests that for the here 
investigated questions it is important to examine the fulfilment of the conditions 
\eqref{eq:34} in each of the considered case. In this sense the following remarks 
are of direct interest.

\textbf{\underline{R.11}}: In the cases described by the wave functions \eqref{eq:6} and 
\eqref{eq:17} for $L _z -\varphi$ and $N - \phi$ pairs one finds 
\begin{equation}\label{eq:37}
\left( {\hat L_z \psi _m ,\,\hat \varphi \psi _m } \right) = \left( 
{\psi _m ,\,\hat L_z \hat \varphi \psi _m } \right) + i\hbar 
\end{equation}
\begin{equation}\label{eq:38}
\left( {\hat N\psi _N \left( \phi  \right),\,\hat \phi \psi _N \left( \phi  
\right)} \right) = \left( {\psi _N \left( \phi  \right),\,\hat N\hat \phi 
\psi _N \left( \phi  \right)} \right)- i
\end{equation}

\textbf{\underline{R.12}}: For $L _z - \varphi$ pair in the cases associated with the 
wave functions \eqref{eq:19} and \eqref{eq:21} one obtains
\begin{equation}\label{eq:39}
\left( {\hat L_z \psi _N \left( \varphi  \right),\,\hat \varphi \psi _N \left( 
\varphi  \right)} \right)= \left( {\psi _N \left( \varphi  \right),\,\hat L_z 
\hat \varphi \psi _N \left( \varphi  \right)} \right)
\end{equation}
\begin{equation}\label{eq:40}
\begin{split}
&\left( {\hat L_z \psi _l ,\,\hat \varphi \psi _l } \right) = \left(
{\psi _l ,\,\hat L_z \hat \varphi \psi _l } \right) +  \\ 
&+ i\hbar \left\{ 1+ 2\, \mathrm{Im} \left[ 
{\sum\limits_{m =  - l}^l {\sum\limits_{r =  - l}^l {c_m^* \, c_r \,\hbar m\left(
{Y_{lm} ,\,\hat \varphi \,Y_{lr} }\right)} } } \right] \right\} 
\end{split}
\end{equation}
(where Im $[\alpha]$ denotes the imaginary part of $\alpha$).

\textbf{\underline{R.13}}: For any wave function $\psi (\varphi)$ with 
$\varphi \in [0, 2 \pi )$ and $\psi (2 \pi - 0) = \psi (0)$ the following relation 
is generally true 
\begin{equation}\label{eq:41}
\left| {\left( {\delta _\psi  \hat L_z \,\psi ,\,\delta _\psi  \hat \varphi \,\psi } 
\right)} \right| \ge \frac{\hbar }{2} \left| {1 - 2 \pi \left| {\psi 
\left( {2 \pi  - 0} \right)} \right|} \right|
\end{equation}

\textbf{\underline{R.14}}: In the case of energy and time  described by the operators
\eqref{eq:24} for any wave function $\psi$ one finds
\begin{equation}\label{eq:42}
\left( {\hat E \psi ,\,\hat t \psi } \right) =\left( {\psi ,\,\hat E\,
\hat t \psi } \right)- i\hbar 
\end{equation}

The things mentioned above in this section justify the next remarks

\textbf{\underline{R.15}}: The Cauchy-Schwarz formula \eqref{eq:33} is an ab origine element in
respect with the usual UR \eqref{eq:2}/\eqref{eq:36}. Moreover, \eqref{eq:33} is always
valid, independently if the conditions \eqref{eq:34} are fulfilled or no. 

\textbf{\underline{R.16}}: The usual UR \eqref{eq:2}/\eqref{eq:36} are valid only in the
circumstances strictly delimited by the conditions \eqref{eq:34} and they are false 
in all other situations.

\textbf{\underline{R.17}}: Due to the relations \eqref{eq:37} and \eqref{eq:38} in the 
cases described by the wave functions \eqref{eq:6} or \eqref{eq:17} the conditions
\eqref{eq:34} are not fulfilled. Consequently in such cases the usual UR 
\eqref{eq:2}/\eqref{eq:36} are essentially inapplicable for the pairs $L _z - \varphi$ respectively
$N - \phi$. However one can see that in the respective cases  the Cauchy-Schwarz  formula
\eqref{eq:33} remains valid as a trivial equality $0 = 0$.

\textbf{\underline{R.18}}: In the cases of EXR described by \eqref{eq:19} the 
$L _z - \varphi$ pair satisfies the conditions \eqref{eq:34} (mainly due to the relation
\eqref{eq:39}). Therefore in the respective cases the usual UR \eqref{eq:2}/\eqref{eq:36}
are valid for $L _z$ and $\varphi$.

\textbf{\underline{R.19}}: The fulfilment of the conditions \eqref{eq:34}  by the 
$L _z -\varphi$ pair for the EXR associated with \eqref{eq:21} depends on the annulment
of the right hand term in \eqref{eq:40} (i.e. on the values of the coefficients $c _m$).
Adequately the correctness of the corresponding UR \eqref{eq:2}/\eqref{eq:36} shows the same
dependence.

\textbf{\underline{R.20}}: The result \eqref{eq:41} points out the fact that the adjusted
relation \eqref{eq:10} is only a secondary piece derivable fom the generally valid Cauchy-Schwarz
formula \eqref{eq:33}.

\textbf{\underline{R.21}}: As regards the energy-time pair the relation \eqref{eq:42} shows
 that the condition \eqref{eq:34} is never satisfied. Consequently for the respective pair
 the UR \eqref{eq:2}/\eqref{eq:36} is not applicable at all. For the same pair, described 
 by the operators \eqref{eq:24}, the Cauchy-Schwarz formula \eqref{eq:33} is always true. But
 because in QM the time $t$ is a deterministic (i.e. non-stochastic) variable in all cases the
 mentioned formula degenerates into the trivial equality $0 = 0$.

 Based on the main notifications from \textbf{R.15 - 20} now we add the following 
 complementary remarks.
 
\textbf{ \underline{R.22}}: For the $L _z - \varphi$ pair the relations \eqref{eq:3} - 
\eqref{eq:4} are always viable in respect with the general Cauchy-Schwarz formula \eqref{eq:33}.
That is why, for a correct description of questions regarding the respective pair, it is
not at all necessarily to replace the mentioned relations with some adjusted formula
like \cite{37}
\begin{equation}\label{eq:43}
\hat L_z = - i\hbar \frac{\partial }{{\partial \varphi }} + \alpha 
\end{equation}
or \cite{14}
\begin{equation}\label{eq:44}
\left[ {\hat L_z ,\,\hat \varphi } \right]= - i\hbar \left( {1 - 2\pi
 \,\delta \left( \varphi  \right)} \right)
\end{equation}
with $\alpha$ = an adjusting constant and $\delta (\varphi)$ = Dirac's function of 
$\varphi$. (in \eqref{eq:44} the notations were adapted to the stipulations 
$\varphi \in [0, 2 \pi )$,  $\hat L_z =  - i\hbar \frac{\partial }{{\partial \varphi }}$
and $\hbar \neq  1$ used in this paper).

Note that the alluded adjustments regard the cases with SCR described by the wave functions \eqref{eq:6} when $\varphi$ plays the role of polar coordinate. 
But for such a role \cite{38}
in order to be a unique (univocal) variable $\varphi$ must be defined naturally only in
the range $[0, 2 \pi)$. (The same range is considered in practice for the normalization of
the wave functions \eqref{eq:6}). Therefore, in the cases under discussion the derivative 
with respect to $\varphi$ refers to the mentioned  range. Particularly for the extremities
of the interval $[0, 2 \pi)$ it has to operate with backward respectively forward 
derivatives. So in the alluded SCR cases the relations \eqref{eq:3} and \eqref{eq:4} act 
well, with a natural correctness. The same correctness is shown by the respective relations
in connection with the EXR described by the wave functions \eqref{eq:19} or \eqref{eq:21}.
In fact, from a more general perspective, the relations \eqref{eq:3} and \eqref{eq:4}
regard the QM operators $\hat L _z$ and $\hat \varphi$. Therefore they must have unique
forms - i.e. expressions which do not depend on the particularities of the considered 
situations (e.g. systems with SCR or with EXR).

\textbf{\underline{R.23}}: The troubles of UR \eqref{eq:2} regarding $L _z - \varphi$ and
$N - \phi$ pairs are directly connected with the conditions \eqref{eq:34}. Then it is 
strange that in the almost all of QM literature the respective conditions are not 
approached adequately. The reason seems to be related with  the nowadays dominant Dirac's
$<bra|$ and $|ket>$ notations. In the respective notations  the terms from the both sides 
of \eqref{eq:34} have a unique representation namely 
$ < \psi |\hat A_j \,\hat A_k |\psi  > $. Such a uniqueness can entail confusion 
(unjustified supposition) that the alluded conditions are always fulfiled. It is 
interesting to note that systematic investigations on the confusions/surprises generated
by the Dirac's notations were started only recently \cite{39}. Probably that further 
efforts on the line of such investigations will bring a new light on the conditions
\eqref{eq:34} as well as on other QM questions.

The findings from the present section give solid arguments for the following concluding
remarks.

\textbf{\underline{R.24}}: In respect with the conjugated observables $L _z - \varphi$,
$N - \phi$ and $E - t$ the usual UR \eqref{eq:2}/\eqref{eq:36} is not adequate for the 
role of normality standard. For such a role the Cauchy-Schwarz
formula \eqref{eq:33} is the most suitable. 
In some cases of interest  the respective formula degenerates in the trivial equality $0 = 0$.

\textbf{\underline{R.25}}: In reality the usual procedures of QM ( illustrated by the 
relations \eqref{eq:3}, \eqref{eq:4}, \eqref{eq:12}, \eqref{eq:15}, \eqref{eq:24} and 
\eqref{eq:25}) work
well and without anomalies in all situations regarding the above mentioned pairs of
observables. Consequently in respect with the conceptual as well as practical interests 
of science the adjusted UR like \eqref{eq:8}, \eqref{eq:9}, \eqref{eq:10} or \eqref{eq:28}
 appear as useless inventions.

\textbf{\underline{R.26}}: Conjointly the previous two remarks prove the fact that the 
cases of the pairs $L _z - \varphi$, $N - \phi$ and $E - t$ infringe in an irrefutable
manner the idea \textbf{\emph{BI.4}} of CIUR. But such a fact must be notified as an 
unsurmountable shortcoming of CIUR doctrine. 

\section{A class of additional shortcomings}\label{sec6}
In the main the CIUR shortcomings discussed above in connection with the pairs
$L _z - \varphi$,  $N - \phi$ and $E - t$ regard the idea \textbf{\emph{BI.4}}. Besides 
this fact the other CIUR ideas, namely \textbf{\emph{BI.1 - 3}} and 
\textbf{ \emph{BI.5}} are troubled by additional shortcomings less reported in 
literature. As a rule, in publications, the respective shortcomings are either ignored
or mentioned on rare occasions. Moreover, even in the alluded occasions the things are
presented separately but not grouped together in reunions destined for collective 
confrontings with CIUR. Here we attempt to put forward such a reunion.

In this attempt we focus our attention on a category of facts able to offer evidences 
about the mentioned additional shortcomings. The announced facts are discussed piece by
piece in the following remarks.

\textbf{\underline{R.27}}: First of all we note the fact that the relation \eqref{eq:1}
is improper for a reference/standard element of a supposed solid doctrine such as CIUR. 
This happens because the respective relations have a transitory/temporary character since
they were founded on old resolution criteria (introduced by Abe and Rayleigh - see 
\cite{30,40}). But the respective criteria were improved in the so-called 
super-resolution techniques  worked out in modern experimental physics \cite{41,42,43,44,45,46,47,48}. 
Then it is possible to imagine some super-resolution-thought-experiments 
($srte$). So, for the corresponding $srte$-uncertainties $\Delta _{srte}A$   and
$\Delta _{srte} B$   of two observables $A$ and $B$ the following relation can be promoted 
\begin{equation}\label{eq:45} 
\Delta _{srte} A \cdot \Delta _{srte} B < \hbar 
\end{equation}
Such a relation is able to replace the CIUR basic formula \eqref{eq:1}. But the alluded
possibility invalidate the idea \textbf{\emph{BI.1}} and incriminate CIUR in connection
with one of its main points.

\textbf{\underline{R.28}}: Secondly let us refer to the term \emph{uncertainty}  used by
CIUR for quantities like $\Delta _\psi A$ from \eqref{eq:2}. We think that the respective 
term is groundless because of the following considerations. As it is defined in the 
mathematical framework of QM (see the previous section) $\Delta _\psi A$ signifies 
the standard deviation 
of the observable $A$ regarded as a stochastic variable. The mentioned framework deals with 
theoretical concepts and models about the intrinsic (inner) properties of the considered 
system but not with elements which refer to the measurements performed on the respective
system. Consequently, for a physical system, $\Delta _\psi A$ refers to the intrinsic
characteristics (reflected in fluctuations) of the observable  $A$. Moreover, the expressions
\eqref{eq:7}, \eqref{eq:20}, \eqref{eq:22} and \eqref{eq:23} reveal the following 
realities: 

(i) for a system in a given state the quantity  $\Delta _\psi A$  has a well 
defined value connected with the corresponding wave function, \\
(ii) the respective value of
$\Delta _\psi A$  is not related with the possible modifications of the  
accuracies regarding the measurement of the observable $A$. 

The alluded realities are 
attested by the fact that for the same state of the measured system (i.e. for the same
value of  $\Delta _\psi A$ ) the measuring uncertainties (regarding $A$) can be changed 
through the improving or worsening of experimental devices/procedures. Note  that the
above mentioned realities imply and justify the observation \cite{49} that,for two 
variables $x$ and $p$ of the same system, the usual CIUR statement \emph{"as $\Delta x$ 
approaches zero, $\Delta p$  becomes infinite and vice versa" }is a doubtful
speculation. Finally we can conclude that the ensemble of the things revealed in the 
present remark contradict the ideas \textbf{ \emph{BI.2 - 4}} of CIUR . But such a 
conclusion must be reported as a serious shortcoming of CIUR.

\textbf{ \underline{R.29}}: The remark \textbf{ R.27} restricts the basic reference
element of CIUR only to the theoretical UR \eqref{eq:2}. On the other hand, as it was 
pointed in Section 5  the UR \eqref{eq:2} is nothing but a secondary and problematical
piece derived from the primary Cauchy-Schwarz formula \eqref{eq:33}. Then it results 
that in discussions about CIUR the piece  UR \eqref{eq:2} must be referred only through its 
affiliation  to the Cauchy-Schwarz formula. Such an affiliation discloses shortcomings 
of CIUR in respect with the pairs of non-commutable  observables $L _z - \varphi$, $N - \phi$ and
$E - t$ (see sections 3, 4 and 5). But note that the mentioned affiliation reveals
shortcomings of CIUR even in the cases of commutable observables . An example of such 
a case one finds with the cartesian momenta $p _x$ and $p _y$ for a particle in a 2D 
potential  well. The well is delimited as follows: the potential energy $V$ is null for 
$0 < x _1 < a$ and $0 < y _1 < b$ respectively $V = \infty$ otherwise, where $0 < a < b$, 
$x_1 = \left( {x + y} \right)/\sqrt 2 $ and $y_1 = \left( {y - x} \right)/
\sqrt 2 $ . For the particle in  the lowest energetic state one finds
\begin{equation}\label{eq:46}
\Delta _\psi  p_x  = \Delta _\psi  p_y  = \hbar \frac{\pi }{{ab}}
\sqrt {\frac{{a^2  + b^2 }}{2}} 
\end{equation}
\begin{equation}\label{eq:47}
\left| {\left\langle {\left( {\delta _\psi  \hat p_x \psi , \, \delta _\psi  
\hat p_y \psi } \right)} \right\rangle } \right| = \left( 
{\frac{{\hbar \pi }}{{ab}}} \right)^2  \cdot \left( {\frac{{b^2  - a^2 }}{2}} \right)
\end{equation}

With these expressions it results directly that for the considered example the momenta
$p _x$ and $p _y$ satisfy the Cauchy-Schwarz formula \eqref{eq:33} in a non-trivial form (i.e. as 
an inequality with a non-null value for the right hand side term). But such a result 
conflicts with the idea \textbf{\emph{BI.4}} and consequently it must be reported as an 
element which incriminates the CIUR doctrine.

\textbf{\underline{R.30}} : The UR \eqref{eq:2}/\eqref{eq:36} is only a particular  
(two-observable) version of the more general many-observable formulae \eqref{eq:31}.
Then for the respective formulae CIUR has to find an interpretation concordant with its
own doctrine (summarized in \textbf{\emph{BI.1 - 5}}). Such an interpretation was proposed
in \cite{50} but it remained as an unconvincing thing (because of the lack of  real 
physical justifications). Other discussions about the formulae \eqref{eq:31} as in 
\cite{13} elude any interpretation of the mentioned kind. A recent attempt \cite{51}
meant to promote an interpretation of relations  like \eqref{eq:31}, for three or more 
observables. But the respective attempt has not a helping value for CIUR doctrine. This is  
because instead of consolidating the CIUR ideas \textbf{\emph{BI.1 - 5}}
it seems  rather to support the idea that the considered relations  are fluctuations 
formulae (in the sense discussed above and bellow in \textbf{ R.28} respectively in
\textbf{R.33} ). We opine that to find a CIUR-concordant interpretation for the 
many-observable formulae \eqref{eq:31} is a difficult (even impossible) task on natural ways
(i.e. without esoteric and/or non-physical considerations). An exemplification of the 
respective difficulty can be appreciated by investigating the case of observables
$A _1 = L _z$, $A _2 = \varphi$ and $A _3 = H$ = energy in the situations described by
the wave functions \eqref{eq:6}, \eqref{eq:19} or \eqref{eq:21}.

\textbf{\underline{R.31}}: The UR \eqref{eq:2}/\eqref{eq:36} fails in the case of 
eigenstates. The fact was mentioned in \cite{52} but it seems to remain unremarked in 
the subsequent publications. In terms of the here developed investigations the alluded 
failure can be discussed as follows. For two non-commutable observables $A$ and $B$ in 
an eigenstate of $A$ one obtains the set of values: $\Delta _\psi A = 0$,
$0 < \Delta _\psi B < \infty$  and 
$\left\langle {\left[ {\hat A,\,\hat B} \right]} \right\rangle _\psi  \ne 0$. But,
evidently, the respective values infringe the UR \eqref{eq:2}/\eqref{eq:36}. Such 
situations one finds particularly with the pairs $L _z - \varphi$ and $N - \phi$ in cases
of states described by the wave functions \eqref{eq:6} and  \eqref{eq:17} respectively
(see the Section 3). A similar example is given by the pair 
$A= H = L_z^2/2I$ = Hamiltonian (energy) and $B = \varphi$ in
respect with states described by the wave functions \eqref{eq:6}.

Now one can see that the question of eigenstates does not engender any problem if the
quantities $\Delta _\psi A$ and $\Delta _\psi B$ are regarded as fluctuations 
characteristics (see the remarks \textbf{R.28} and \textbf{R.33}). Then the mentioned set
of values show that in the respective eigenstate $A$ has not fluctuations (i.e. $A$
behaves as a deterministic variable) while $B$ is endowed with fluctuations (i.e. $B$
appears as a stochastic variable). Note also that in the cases of specified eigenstates 
the UR \eqref{eq:6}/\eqref{eq:36} are not valid. This happens because of the fact that 
in such cases the conditions \eqref{eq:34} are not satisfied. The respective fact 
is proved by the observation that its opposite imply the absurd result
\begin{equation}\label{eq:48}
a \cdot \left\langle B \right\rangle _\psi  = \left\langle {\left[ {\hat A,\,
\hat B} \right]} \right\rangle _\psi   + a \cdot \left\langle B \right\rangle _\psi  
\end{equation}
with $\left\langle {\left[ {\hat A,\,\hat B} \right]} \right\rangle _\psi  \ne 0$ 
and $a$ = eigenvalue of ${\hat A}$ (i.e. ${\hat A} \psi = a \psi$  ). But in the cases 
of the alluded eigenststes the Cauchy-Schwarz formula \eqref{eq:33} remain valid. 
It degenerates into the trivial equality $0 = 0$ (because $\delta _\psi  \hat A\,\psi  = 0$). So
one finds a contradiction with \textbf{\emph{BI.4}} - i.e. an additional and distinct
shortcoming of CIUR.

\textbf{\underline{R.32}}: Now let us note the fact UR \eqref{eq:2}/\eqref{eq:36} as well
as the relations \eqref{eq:31} and \eqref{eq:33} are one-temporal  formulae because all the
implied quantities refer to the same instant of time. But the mentioned formulae can be
generalized into multi-temporal versions, in which the corresponding  quantities refer to
different instants of time. So \eqref{eq:33} is generalizable in the form
\begin{equation}\label{eq:49}
\Delta _{\psi _1 } A \cdot \Delta _{\psi _2 } B \ge \left| {\left( 
{\delta _{\psi _1 } \hat A\,\psi _1 ,\,\delta _{\psi _2 } \hat B \psi _2 } 
\right)} \right|
\end{equation}
where $\psi _1$ and $\psi _2$ represent the wave function for two different instants of
time $t _1$ and $t _2$. If in \eqref{eq:49} one takes $|t _2 - t _1 | \rightarrow \infty$
in the CIUR vision the quantities $\Delta _{\psi _1 } A\,$ and $\Delta _{\psi _2 } B$ 
have to refer to $A$ and $B$ regarded as independent solitary observables. But in such a
regard if $\left( {\delta _{\psi _1 } \hat A\,\psi _1 ,\,\delta _{\psi _2 } 
\hat B\,\psi _2 } \right) \ne 0$ the relation \eqref{eq:49} refute the idea 
\textbf{\emph{BI.2}} and so it reveals  another additional shortcoming of CIUR. Note
here our opinion that the various attempts \cite{53}, \cite{54} of extrapolating the
CIUR vision onto the relations of type \eqref{eq:49} are nothing but artifacts without
any real (physical) justification. We think that the relation \eqref{eq:49} does not
engender  any problem if it is regarded as fluctuations formula (in the sense mentioned
in \textbf{R.28} and \textbf{R.33}). In such a regard the cases when $\left( {\delta _
{\psi _1 } \hat A\,\psi _1 ,\,\delta _{\psi _2 } \hat B\,\psi _2 \,} \right) 
\ne 0$ refer to the situations in which, for the instants $t _1$ and $t _2$, the
corresponding fluctuations of $A$ and $B$ are correlated (i.e. statistically dependent).

\textbf{\underline{R.33}}: Now let us call attention on a quantum-classical 
similarity which directly contradicts  the  idea \textbf{\emph{BI.5}} of CIUR. The respective similarity directly
regards  the UR \eqref{eq:2}/\eqref{eq:36} as descendant from the relations
\eqref{eq:33} and \eqref{eq:31}. Indeed the mentioned relations are completely analogous 
with a set of classical formulas from phenomenolgical theory of fluctuations. The alluded
formulae can be written \cite{55}, \cite{56} as follows
\begin{equation}\label{eq:50}
\det \left[ {\left\langle {\delta _w {\rm A}_j \,\delta _w {\rm A}_k } 
\right\rangle _w } \right] \ge 0
\end{equation}
\begin{equation}\label{eq:51}
\Delta _w {\rm A} \cdot \Delta _w {\rm B} \ge \left| {\left\langle 
{\delta _w {\rm A}\,\delta _w {\rm B}} \right\rangle _w } \right|
\end{equation}
In these  formulae  ${{\rm A}_j }$ , ${\rm A}$ and ${\rm B}$ signify 
the classical global observables which characterize a thermodynamic system in its wholeness.
In the same formulae $w$ denotes the phenomenological probability distribution, 
$\left\langle \ldots \right\rangle _w $ represents the mean (expected value) evaluated 
by means of $w$  while $\Delta _w {\rm A}$ , $\Delta _w {\rm B}$ and
${\left\langle {\delta _w {\rm A}\,\delta _w {\rm B}} \right\rangle _w }$ stand for 
characteristics (standard deviations respectively correlation) regarding the fluctuations of the mentioned
observables. We remind the appreciation that in classical physics the alluded 
characteristics and, consequently, the relations \eqref{eq:50} - \eqref{eq:51} describe 
the intrinsic (own) properties of thermodynamic systems but not the aspects of measurements
performed on the respective systems. Such an appreciation is legitimated for example by 
the research regarding the fluctuation spectroscopy \cite{57} where the properties of 
macroscopic (thermodynamic) systems are evaluated through the (spectral components of)
characteristics like $\Delta _w {\rm A}$ and ${\left\langle {\delta _w {\rm A}\,\delta _w {\rm B}} \right\rangle _w }$ .

The above discussions disclose the groundlessness of  idea \cite{58}-\cite{60}  that
the relations like \eqref{eq:51} have to be regarded as a sign of a macroscopic/classical 
complementarity (similar with the quantum complementarity supposed by CIUR idea
\textbf{\emph{BI.4}}). According to the respective idea the quantities
$\Delta _w {\rm A}$ and $\Delta _w {\rm B}$ appear as macroscopic uncertainties. Note that
the mentioned idea was criticized partially in \cite{61,62} but without any explicit 
specification that the quantities $\Delta _w {\rm A}$ and $\Delta _w {\rm B}$ are 
characteristics of fluctuations. 

The previously notified quantum-classical similarity together with the reminded 
significance of the quantities implied in \eqref{eq:50} and \eqref{eq:51} suggests and consolidates the 
following regard (argued also in \textbf{R.28}). The quantities $\Delta _\psi A$ and
$\Delta _\psi B$ from UR \eqref{eq:2}/\eqref{eq:36} must be regarded as  describing 
intrinsic properties (fluctuations) of quantum observables $A$ and $B$ but not as uncertainties of 
such observables.

Now, in conclusion, one can say that the existence of classical relations \eqref{eq:50}
and \eqref{eq:51} contravenes to both ideas \textbf{\emph{BI.5}} and \textbf{\emph{BI.1}} 
of CIUR.

\textbf{\underline{R.34}}: In classical physics the fluctuations of ${\rm A}$ and 
${\rm B}$ implied in \eqref{eq:51} are described not only by the second order parameters 
like $\Delta _w {\rm A}$ , $\Delta _w {\rm B}$ or ${\left\langle {\delta _w {\rm A}\,
\delta _w {\rm B}} \right\rangle _w }$. For a better evaluation the respective fluctuations
are characterized additionally \cite{63} by higher order moments like
$\left\langle {\left( {\delta _w {\rm A}} \right)^r \,\left( {\delta _w
{\rm B}} \right)^s } \right\rangle _w $ with $r + s \geq 3$. This fact suggests the
observation that, in the context considered by CIUR, we also have to use  the quantum higher 
order moments like $\left\langle {\left( {\delta _\psi  A} \right)^r \,\left( 
{\delta _\psi  B} \right)^s } \right\rangle _\psi $ with $r + s \geq 3$. Then for the 
respective quantum moments CIUR is obliged to offer an interpretation compatible with
its own doctrine. But it seems to be less probable that such an interpretation can be 
promoted through credible (and natural) arguments.

\textbf{ \underline{R.35}}: The thermodynamic systems were also implied  in other debates
about CIUR, in connection with the question of the so called "macroscopic operators"
(see \cite{64}, \cite{65} and references). The question appeared as follows. By analogy with UR
\eqref{eq:2}  and viewing the respective systems in terms of quantum statistical
physics, the CIUR partisans promoted the formula
\begin{equation}\label{eq:52}
\Delta _\rho  {\rm A} \cdot \Delta _{\rho } {\rm B} \ge \frac{1}{2} \left|
{\left\langle {\left[ {\hat {\rm A},\,\hat {\rm B}} \right]} \right\rangle _\rho  } 
\right|
\end{equation}
This formula implies the notations: ${\rm A}$ and ${\rm B}$ denote two global observables 
(for the system in its wholeness) described by the corresponding operators
${\hat {\rm A}}$ and ${\hat {\rm B}}$,  $\hat\rho$ signifies the statistical operator
(density matrix) associated with the global state of the system, respectively
$\Delta _\rho  {\rm A} = \left\{ {Tr\,\left[ {\left( {\hat {\rm A} - 
\left\langle {\rm A} \right\rangle _\rho  } \right)^2 \,\hat \rho } \right]} \right\}^
{\frac{1}{2}} $ where $\left\langle {\rm A} \right\rangle _\rho   = Tr\left(
{\hat {\rm A}\,\,\hat \rho } \right)$ = the mean value of ${\rm A}$. Relation \eqref{eq:52}
entailed discussions because of the conflict between the following two findings:

(i) On  one hand \eqref{eq:52} is introduced by analogy with UR \eqref{eq:2}/\eqref{eq:36}
on which CIUR is founded. Then, by extrapolating CIUR, the quantities 
$\Delta _\rho  {\rm A}$ and $\Delta _\rho  {\rm B}$ from \eqref{eq:52}  should be 
interpreted as (global) uncertainties subjected to stipulations as the ones indicated in 
\textbf{\emph{BI.3}} and \textbf{\emph{BI.4}}.\\
(ii) On the other hand, in the spirit of \textbf{\emph{BI.5}}, CIUR agrees with the idea 
that  the observables characterizing the thermodynamic systems are possible to be
measured without any uncertainty (i.e. with unbounded accuracy). For an observable the
mentioned  possibility should be independent of the fact that it is measured solitarily or 
simultaneously with other observables. Thus, for two thermodynamic observables, it is 
senselessly to accept stipulations such are the ones prescribed by \textbf{\emph{BI.3}} 
and \textbf{\emph{BI.4}}.

In order to elude the mentioned conflict a strange purpose was promoted, namely: to 
abrogate the formula \eqref{eq:52} and to replace it with an adjusted macroscopic 
relation concordant with CIUR vision.  For such a purpose the global operators 
${\hat {\rm A}}$ and ${\hat {\rm B}}$  from \eqref{eq:52} were substituted \cite{64}, \cite{65}
by the so-called "macroscopic operators" $\hat{\mathcal{A}}$ and ${\hat{\mathcal B}}$.
The respective "macroscopic operators"  are considered to be representable as
 quasi-diagonal matrices (i.e. as matrices with non-null elements only in a 
"microscopic neighbourhood" of principal diagonals). Then one supposes  that
$\left[ \hat{\mathcal A},\,\hat{\mathcal B} \right] = 0$ for any pairs of
observables ${\rm A}$ and ${\rm B}$ and, consequently instead of \eqref{eq:52} one obtains
\begin{equation}\label{eq:53}
\Delta _\rho  \hat {\mathcal A} \cdot \Delta _\rho  \hat {\mathcal B} \ge 0
\end{equation}
In this formula CIUR partisans see the fact that the uncertainties $\Delta _\rho  {\rm A}$
and $\Delta _\rho  {\rm B}$ can be unboundedly small at the same instant of time. Such a
fact is in concordance with CIUR  vision about macroscopic observables. Today it seems 
to be accepted the belief that the adjusted relation \eqref{eq:53} solves all the troubles 
of CIUR caused by the formula \eqref{eq:52}.

A first disapproval of the mentioned belief results from the following observations:\\
(i) Relation \eqref{eq:52} cannot be abrogated if the entire mathematical apparatus of 
quantum statistical physics is not abrogated too. More exactly, the substitution of
operators from the global version $\hat {\rm A}_j $ into a "macroscopic" variant 
$\hat {\mathcal A}_j$   is a senseless invention as long as in practical procedures
of quantum statistical physics \cite{66}, \cite{67} as lucrative operators one uses 
$\hat {\rm A}_j $ but not  $\hat {\mathcal A}_j$.\\
(ii) The substitution $\hat {\rm A}_j $  $\rightarrow$  $\hat {\mathcal A}_j$ does not 
metamorphose automatically \eqref{eq:52} into \eqref{eq:53},  because if two 
operators are quasi-diagonal, in  sense required by the partisans of CIUR, it is not 
surely that they commute.  As an example  we quote the Cartesian components
of the global magnetization ${\vec {\rm M}}$ of a paramagnetic system formed of 
$\emph{N}$ independent $\frac{1}{2}$ - spins. The alluded components are described by
the global operators
\begin{equation}\label{eq:54}
\hat {\rm M}_\alpha  = \frac{{\gamma \hbar }}{2}\,\hat \sigma _\alpha ^{(1)} 
\, \oplus \,\frac{{\gamma \hbar }}{2}\,\hat \sigma _\alpha ^{(2)} \, \oplus \ldots 
\oplus \,\frac{{\gamma \hbar }}{2}\,\hat \sigma _\alpha ^{(N)} 
\end{equation}
where $\alpha = x, y, z$ ;  $\gamma$ = magneto-mechanical factor and 
$\,\hat \sigma _\alpha ^{(i)} $ = Pauli matrices associated to the $i$-th spin (particle).
Note that the operators \eqref{eq:54} are quasi-diagonal in the mentioned sense but, 
for all that, they do not commute because
$\left[ {\hat {\rm M}_\alpha  \,,\;\hat {\rm M}_\beta  } \right]\,\; = \,\;i \hbar 
\gamma \,\epsilon _{\alpha \beta \mu } \,\hat {\rm M}_\mu $ 
($\epsilon _{\alpha \beta \mu } $ denote the Levi-Civita tensor).

A second disproval of the belief induced by the substitution 
$\hat {\rm A}_j $  $\rightarrow$  $\hat {\mathcal A}_j$ is evidenced if the
relation \eqref{eq:52} is regarded in an ab origine (and natural) approach similar 
with the one presented in Section 5. In such  regard it is easy to see that in fact 
the formula \eqref{eq:52} is only a restrictive descendant from the generally valid
relation
\begin{equation}\label{eq:55}
\Delta _\rho  {\rm A} \cdot \Delta _\rho  {\rm B}\,\;\; \ge \;\;\,\left| {\left\langle 
{\delta _\rho  \hat {\rm A}\,\delta _\rho  \hat {\rm B}} \right\rangle _\rho  } \right|
\end{equation}
where $\delta _\rho  \,\hat {\rm A}\,\; = \,\;\hat {\rm A}\, - \,\left\langle {\rm A} 
\right\rangle _\rho  $.

This last relation justifies the following affirmations:

(i) Even in the situations when $\left[ {\hat {\rm A}\,,\;\hat {\rm B}}
\right]\,\; = \,\;0$ the product $\Delta _\rho  {\rm A} \cdot \Delta _\rho  {\rm B}$
can be lower bounded by a non-null quantity. This happens because it is possible to 
find cases
in which the term from the right hand side of (55) has a non-null value.\\
(ii) Relations \eqref{eq:55} remain valid, without any problem, even after the 
substitution  $\hat {\rm A}_j $  $\rightarrow$  $\hat {\mathcal A}_j$. Then according 
to the previous affirmation the respective substitution does not guaranttee the relation 
\eqref{eq:53} and the corresponding speculations.\\
The above presented facts warrant the conclusion that the relation \eqref{eq:52} reveal
a real shortcoming of CIUR. The respective shortcoming cannot be avoided by restoring 
to the so-called "macroscopic operators". But note that the same relation does not
cause any problem if it is considered together with \eqref{eq:55} as formulae which 
refer to the fluctuations of global observables regarding thermodynamic systems.\\    

\textbf{ \underline{R.36}}: The quantum-classical similarity revealed in \textbf{R.33} 
 also entails a proof  against the CIUR assertion from \textbf{\emph{BI.5}} that Planck
constant $\hbar$ has no analog in non-quantum physics. Such a proof results from the 
following facts. The alluded similarity regards the groups of relations \eqref{eq:31},
\eqref{eq:33}, \eqref{eq:36}/\eqref{eq:2} and \eqref{eq:50}, \eqref{eq:51}. The respective
relations imply the standard deviations $\Delta _\psi A _j$ or $\Delta _w {\rm A}_j$
associated with the fluctuations of the corresponding observables. But mathematically 
the standard deviation indicate the stochasticity  (randomness) of a variable, in the 
sense that it has a positive or null value as the respective variable is a stochastic or,
alternatively, deterministic (non-stochastic) quantity.  Therefore the deviations 
$\Delta _\psi A _j$ and $\Delta _w {\rm A}_j$ can be regarded as similar indicators of 
stochasticity for the quantum respectively classical observables.

For diverse cases (of observables, systems and states) the classical deviations have 
various expressions in which, apparently, no common element seems to be  implied.
Nevertheless such an element can be found out \cite{68} as being materialized by the
Boltzmann constant $k _B$. So, in the framework of phenomenological theory of fluctuations
(in Gaussian approximation) one obtains \cite{68}
\begin{equation}\label{eq:56}  
\left( {\Delta _w \,{\rm A}_j } \right)^2  = k_B \,\sum\limits_\alpha  
{\sum\limits_\beta  {\frac{{\partial \bar {\rm A}_j }}{{\partial \bar {\rm X}_
\alpha  }}} } \,\frac{{\partial \bar {\rm A}_j }}{{\partial \bar {\rm X}_\beta  }}\,
\left( {\frac{{\partial ^2 \bar {\rm S}}}{{\partial \bar {\rm X}_\alpha  \,\partial \bar 
{\rm X}_\beta  }}} \right)^{ - 1} 
\end{equation}
In this relation $\bar {\rm A}_j = \left\langle {{\rm A}_j } \right\rangle _w $, 
${\rm S} = {\rm S}\left( {{\rm X}_\alpha  } \right)$ denotes the entropy written as 
a function of independent variables ${{\rm X}_\alpha  }$ , $(\alpha = 1, 2, \ldots, r)$ and
$\left( {a_{\alpha \beta } } \right)^{ - 1} $ represent the elements of the inverse of
matrix $\left[ {a_{\alpha \beta } } \right]$. Then from \eqref{eq:56} it result that the 
expressions for $\left( {\Delta _w \,{\rm A}_j } \right)^2 $ consist of products of $k _B$
with factors which are independent of $k _B$. The respective independence is evidenced by 
the fact that the alluded factors ought to coincide with deterministic (non-stochastic)
quantities from usual thermodynamics. Or it is known that such quantities  do not imply 
$k _B$ at all. Concrete exemplifications of the relations \eqref{eq:56} with the above 
noted properties are quoted in \cite{68}.

Then, as a first aspect, from \eqref{eq:56} it results that the fluctuations characteristics 
(dispersions)  $\left( {\Delta _w \,{\rm A}_j } \right)^2 $ are directly proportional to
$k _B$ and, consequently, they are non-null  respectively null quantities as 
$k _B \neq 0 $ or $k _B \rightarrow 0$. (Note that because $k _B$ is a constant the limit
$k _B \rightarrow 0$ means that the quantities directly proportional with $k _B$ are 
negligible comparatively with other quantities of same dimensionality but independent of 
$k _B$ ). On the other hand, the second aspect (mentioned also  above) is the fact that
$\Delta _w \,{\rm A}_j  $ are particular indicators of classical stochasticity. 
Conjointly the two mentioned aspects show that $k _B$ has the qualities 
of an authentic  generic indicator of thermal stochasticity which is specific for classical 
macroscopic systems. (Add here the observation that the same quality of $k _B$ can be
revealed also \cite{68} if the thermal stochasticity is studied in the framework of
classical statistical mechanics).

Now  let us discuss about the quantum stochasticity whose indicators are the standard 
deviations $\Delta _\psi A _j$. Based on the relations  \eqref{eq:20} and
\eqref{eq:46} one can say that in many cases the expressions for 
$(\Delta _\psi A _j)^2$ consist in products of Planck constant $\hbar$ with factors 
which are independent of $\hbar$. Then, by analogy with the above discussed classical
situations, $\hbar$ places itself in the posture of generic indicator for quantum 
stochasticity. 

In the alluded posture the Planck constant $\hbar$ has an authentic classical analog 
represented by the Boltzmann constant $k _B$. But such an analogy contradicts strongly 
the idea \textbf{\emph{BI.5}}.

In connection with the roles of $k _B$ and $\hbar$ as generic indicators of stochasticity
it is of interest to add here the following aspect. In their above presented 
roles $k _B$ and $\hbar$ regard the one-fold stochasticity, of classical and  
quantum nature respectively, evaluated through the deviations  $\Delta _w \,{\rm A}_j$ 
and $\Delta _\psi A _j$. But in physics is also known  a two-fold stochasticity, of a
combined thermal and quantum nature. Such a stochasticity appears in cases of 
quantum statistical systems and it is evaluated through the standard deviations
$\Delta _\rho  \,{\rm A}_j $ implied in relations \eqref{eq:52} and \eqref{eq:55}.
The expressions of the mentioned deviations can be obtained by means of the 
fluctuation-dissipation theorem \cite{66}. Therefore
\begin{equation}\label{eq:57}
\left( {\Delta _\rho  \,{\rm A}_j } \right)^2  = \frac{\hbar }{{2\pi }}\,
\int\limits_{ - \infty }^\infty  {\coth \left( {\frac{{\hbar \omega }}{{2k_B T}}} 
\right)\,\,} \chi _{jj}^{''} \left( \omega  \right)\, d\omega 
\end{equation}
Here $\chi _{jj}^{''} \left( \omega  \right)$ denote the imaginary parts of
the susceptibilities  associated with the observables ${\rm A} _j$. Note that
$\chi _{jj}^{''} \left( \omega  \right)$ are the deterministic quantities which 
appear also in non-stochastic framework of macroscopic physics \cite{69}. That is why  
$\chi _{jj}^{''} \left( \omega  \right)$ are independent of both $k _B$ and $\hbar$. Then
from \eqref{eq:57} it results that $k _B$ and $\hbar$ considered together appear as
a couple of generic indicators for the twofold stochasticity of thermal and quantum 
nature. The respective stochasticity is negligible when $k _B \rightarrow 0$ and
$\hbar \rightarrow 0$ and  significant when $k _B \neq 0$ and $\hbar \neq 0$ 
respectively. 

The above discussions about the quantum stochasticity and the limit $\hbar \rightarrow 0$
must be supplemented with the following specifications. The respective stochasticity
regards the cases of observables of orbital and  spin types respectively. In the 
orbital cases the limit $\hbar \rightarrow 0$ is usually associated  with the quantum$\rightarrow$
classical limit. The respective limit implies an unbounded growth of the values of some
quantum numbers so as to ensure a correct limit for the orbital movements. Then one finds
\cite{70,71} that the orbital-type stochasticity is in one of the following two 
situations:

(i) in the mentioned limit it converts oneself in a classical-type stochasticity of the
corresponding observables (e.g. in the cases of $\varphi$ and $L _z$ of a torsional 
pendulum or of $x$ and $p$ of a rectilinear oscillator), or\\
(ii) in the same limit it disappears, the corresponding observables  becoming deterministic 
classical variables (e.g. in the case of the distance $r$ of the electron in respect 
with the nucleus in a hydrogen atom).

The quantum stochasticity of spin-type regards the spin observables. In the limit 
$\hbar \rightarrow 0$  such observables disappear completely (i.e. they lose both their 
mean values and the affined fluctuations).

We end here this section where the remarks \textbf{R.27 - 36} point out a class of less
discussed shortcomings of CIUR. The respective class supplements the set of the more known 
 of CIUR defects discussed in sections \ref{sec3}, \ref{sec4}, and \ref{sec5}. In its wholeness the resulting set 
 contradicts indubitably  the ensemble of all CIUR basic ideas \textbf{\emph{BI.1 - 5}}.

\section{A first reconsideration of the things: Abandonment of CIUR}\label{sec7}
In sections 3 - 6 we have presented a set of shortcomings whose ensemble contradicts in 
an indubitable manner all the basic ideas \textbf{\emph{BI.1 - 5}} of CIUR. Of course, the 
respective presentation ought to be supplemented with specifications regarding both the
gravity of the things and the possible reconsideration  of the discussed  questions. In
this sense we note the following remarks.

\textbf{\underline{R.37}}: The mentioned contradictions are irrefutable for CIUR doctrine 
in the sense that they  can not be surmounted by inner arguments (deductible from 
\textbf{\emph{BI.1 - 5}}) of the respective doctrine. Consequently the vexed question 
about CIUR must be approached constructively by taking into account the alluded 
shortcomings and by looking for adequate reappraisal of the facts.

\textbf{\underline{R.38}}: In the mentioned circumstances CIUR proves oneself to be an 
incorrect amalgam (of suppositions) deprived of necessary qualities of a valid scientific
construction. That is why CIUR must be abandoned as a wrong doctrine which, in fact, has 
no real value.

\textbf{\underline{R.39}}: The alluded abandonment has to be completed by a natural 
re-interpretation of the basic CIUR's relations \eqref{eq:1} and \eqref{eq:2}. We opine 
that the respective re-interpretation is argued mainly in the remarks \textbf{R.27}, 
 \textbf{R.28} and \textbf{R.33} from Section 6. So the relations \eqref{eq:1} appear 
as fictions without any physical significance. On the other hand the relations 
\eqref{eq:2} are simple fluctuations formulae, from the same family with the 
microscopic and macroscopic relations from the groups \eqref{eq:31}, \eqref{eq:33} and
\eqref{eq:36} respectively \eqref{eq:50}, \eqref{eq:51} and \eqref{eq:55}. Consequently 
the relations \eqref{eq:1} and \eqref{eq:2} have no special or extraordinary 
status/significance in physics.

\textbf{\underline{R.40}}: The reappraisals noted in \textbf{R.38} and \textbf{ R.39} do
not disturb in any way the framework (conceptions and procedures) of usual QM as it is
applied concretely in the   investigations of quantum systems.

\textbf{\underline{R.41}}: The above alluded reappraisals  disconnect the relations
\eqref{eq:1} and \eqref{eq:2} from the description of quantum measurements. So the 
respective description become a distinct scientific question. It will be discussed in
the next sections.

\section{An inspection of the conventional views about quantum measurements}\label{sec8}
The question regarding the description of the quantum measurements (QMS) is one of 
the most debated subject associated with the CIUR history. It generated a large diversity of viewpoints relatively
to its importance and/or approach (see \cite{1,25,26,27,28,29} and references). 
The respective diversity inserts even some extreme opinions such are:

(i) the description of QMS is \emph{"probably the most important part of the theory 
(\emph{"QM"})"} \cite{1}.\\
(ii) \emph{"the word (\emph{"measurement"}) has had such a damaging effect on the  
discussions that ... it should be banned altogether in quantum mechanics"} \cite{72}.

As a notable aspect today one finds that the many of the existing approaches of QMS 
(including some of the most recent ones) are of conventional essence. This happens because 
they are grounded on some conventional premises (\textbf{\emph{CP}}) which presume and 
even try to 
extend the CIUR doctrine (see \cite{1,6,25,26,27,28,29,73,74,75} and references). 
That is why, a reconsideration of CIUR like the one 
presented above in the previous sections, requires a corresponding inspection of the 
mentioned \textbf{\emph{CP}} regarding the QMS. We start such an inspection by pointing out the 
fact that, in essence, the alluded \textbf{\emph{CP}} can be resumed as follows:

\textbf{\emph{\underline{CP.1}}}  \emph{(\underline{basic}): The descriptions of QMS 
must be developed as confirmations and  extensions of CIUR doctrine.}

\textbf{\emph{\underline{CP.2}}} \emph{(\underline{supporting CIUR}): The peculiarities of QMS reported in
\textbf{BI.2 -4} are connected with the corresponding features of the measuring 
perturbations (which trouble the investigated systems during the measurements). So in the
cases of observables refered in \textbf{BI.2 - 3} respectively in \textbf{ BI.4 } the 
alluded perturbations are supposed to have an avoidable respectively an unavoidable
character.} 

\textbf{\emph{\underline{CP.3}}} \emph{(\underline{supporting CIUR}): In the case 
of QMS the mentioned 
perturbations cause specific jumps in states of the measured systems. The respective jumps
have to be included obligatory in the descriptions of QMS.}

\textbf{\emph{\underline{CP.4}}} \emph{(\underline{supporting CIUR}): With regard 
to the observables 
of quantum and  classical type respectively the measuring inconveniences (perturbations and 
uncertainties) show an essential difference. Namely they are unavoidable respectively 
avoidable characteristics of measurements. The mentioned difference must be taken into 
account as a main point in the descriptions of the measurements regarding the two types
of observables.} 

\textbf{\emph{\underline{CP.5}}} \emph{\underline{(extending CIUR}): For a 
quantum observable $A$ of a system in the state $| \psi >$ a QMS is assumed to give as result a single value say  
$a _n$ which is one of the eigenvalues of the associated operator $\hat A$. Therefore
the description of the respective QMS must include as essential piece a sudden reduction
(collapse) of the wave function i.e a relation of the form:}
\begin{equation}\label{eq:58}
|\psi (t)> \text{before  measurement}  \rightarrow   |a _n > \text{after measurement}
\end{equation}
\emph{where $t$ denotes the instant of QMS and $|a _n >$ represents the eigenfunction of
$\hat A$ corresponding to the eigenvalue $a _n$.}
\textbf{\emph{\underline{CP.6}}} \emph{(\underline{extending CIUR}): The 
description of QMS ought
to be incorporated as an inseparable part in the framework of QM. Adequately QM must be 
considered as a unitary theory both of intrinsic properties of quantum systems and 
of measurements regarding the respective properties.} 

Now the announced inspection of the conventional  conceptions about QMS  can be focused 
in comments on the above premises \textbf{\emph{CP.1 - 6}}. In the spirit of the previous
discussions regarding CIUR for the alluded comments of prime importance is to note that
the premises \textbf{\emph{CP.1 - 6}} are troubled by many sortcomings. The respective 
troubles are pointed out piece by piece through the following remarks.

\textbf{\underline{R.42}}: As we concluded in Section 7 in fact CIUR is nothing but a wrong
doctrine which must be abandoned. Consequently CIUR has to be omitted from the lucrative 
scientific discussions. Moreover it is illegitimately to approach a scientific question
(as is the description of QMS) by using and extending such a doctrine. That is why the
premise \textbf{\emph{CP.1}} is totally groundless.

\textbf{\underline{R.43}}: The premise \textbf{\emph{CP.2}} is inspired and argued by the 
ideas of CIUR about the relations \eqref{eq:1} and \eqref{eq:2}. But, according to the
discussions from the previous sections, the respective ideas are completely unfounded.
Therefore the alluded \textbf{\emph{CP.2}} is deprived of any necessary and well-grounded
justification.

\textbf{\underline{R.44}}: In the main \textbf{\emph{CP.3}} is inferred from the belief that
the mentioned jumps have an essential importance for QMS. But the respective belief 
appears as entirely unjustified if one takes into account the following natural and 
indubitable observation \cite{76}: \emph{"it seems essential to the notion 
of measurement that it answers  a question about the given situation existing before the measurement. Whether
the measurement leaves the measured system unchanged or brings about a new and different 
state of that system is a second and independent question"}. So we  have to report
an inevitable deficiency of \textbf{\emph{CP.3}}. 

\textbf{\underline{R.45}}: The essence of the difference mentioned in \textbf{\emph{CP.4}}
is questionable at least because of the following two reasons:

(i) In the classical  case the mentioned avoidance of the measuring inconveniences 
have not a significance of principle but only a relative and limited value 
(depending on the performances of measuring devices and procedures). Such a fact seems to be well known 
by experimenters. \\ 
(ii) In the quantum case until now the alluded unavoidableness  cannot be justified 
by valid arguments of experimental nature (see the above remark \textbf{R.27} and the
comments regarding the relation \eqref{eq:45}).

\textbf{\underline{R.46}}: Through the assumption implied in \textbf{\emph{CP.5}} a QMS
is presumed to consist of a single experimental trial. But as it is well known the quantum 
observables are stochastic variables endowed with specific spectra of values. On the 
other hand, from a mathematical perspective \cite{36}, for a stochastic variable a single
experimental trial (outcome) has no significance. A true  measurement (experimental 
evaluation) of such a variable requires a statistical sampling composed by a (large) number of
single trials. The results of the mentioned sampling facilitate the estimation of the
probabilistic characteristics (e.g. mean value and standard deviation) of the respective
variable. The mentioned features of measurements regarding stochastic observables are 
taken into account in the classical (non-quantum) context in connection with the 
phenomenological  theory of fluctuations  \cite{77,78}. In the respective context 
a stochastic observable $\rm A$  is characterized \cite{79,55,56} by a probability 
distribution $w (a)$. But for a true measurement of $\rm A$ 
a single  experimental trial, which gives a unique value - say $a _0$, has no significance. 
Consequently, there  it is no interest for a reduction (collapse) of probability 
distribution like
\begin{equation}\label{eq:59}
w\left( a \right) \quad \text{before measurement} \to \delta \left( {a - a_0 } 
\right) \quad \text{after measurement}
\end{equation} 
with $\delta$ = Dirac function.

Then by a credible analogy one can say that in respect with QMS  the reduction/collapse 
\eqref{eq:58} has not any real scientific meaning.

\textbf{\underline{R.47}}: The reduction/collapse \eqref{eq:58} also appears  as 
meaningless  if the question of QMS is regarded from the perspective of Albertson's 
observation \cite{76} quoted above in \textbf{R.44}.

\textbf{\underline{R.48}}: The premise \textbf{\emph{CP.6}} proves  to be an 
unjustified idea if  the usual conventions of physics are considered. According to the 
respective conventions, in all the basic chapters of physics, each observable of a system is 
regarded as a concept \emph{ "per se" } (in its essence) which is denuded of measuring 
aspects. Or QM is nothing but such a basic chapter, like classical mechanics, 
thermodynamics, electrodynamics or statistical physics. On the other hand in physics the measurements 
appear as main pourposes of experiments. But note that the study of the experiments
has its own problems \cite{80} and  is done in frameworks which are additional
and distinct in respect with the basic chapters of physics. The above note is 
consolidated by the observation that \cite{81}: \emph{"the procedures of measurement 
(comparison with standards) has a part which cannot be described inside the branch of physics where
it is used"}.

Then, in contrast with the premise \textbf{\emph{CP.6}}, it is natural to accept the 
idea that QM and the description of QMS have to remain  distinct scientific branches.
However the two branches have to use some common concepts and symbols. This happens because, 
in fact, both of them also imply  elements regarding the same quantum systems.

In the end of this section in  can be seen that remarks \textbf{R.42 - 48} point out 
serious shortcomings of all  premises \textbf{\emph{CP.1 - 6}} regarding QMS.  Consequently
the conventional approaches of QMS prove themselves to be unsuccesful endeavours. Then, 
with regard to QMS, it can be of some nontrivial interest  to search for possible
new approaches, dissociated from the premises \textbf{\emph{CP.1 - 6}}. Such an approach 
is presented in the next sections. 

\section{A second reconsideration of  things: New views about QMS}\label{sec9}
It is known  that the above presented premises \textbf{\emph{CP.1 - 6}} persist 
in  nowadays publications regarding the problem  of QMS description. Then, by taking into 
account the remarks \textbf{R.42 - 48} from the previous section, it results that the
mentioned problem is still an open question (at least partially). The respective question 
requires reconsidered approaches founded on new premises (\textbf{\emph{NP}}) and 
disconnected of conventional premises \textbf{\emph{CP.1 - 6}} and CIUR doctrine.  We attempt to present such 
\textbf{\emph{NP}} in this section.

Firstly we note that a natural theory of measurements must contain elements (concepts and
reasonings) which are in adequate correspondence with the main characteristics of the
real measuring experiments. Then, for our attempt, it is of major interest to account the 
alluded characteristics of QMS regarded in the general context of scientific practice 
(with its proved and accepted views). We start the mentioned account with the observation 
that in  classical physics the belief \cite{82} \emph{"in the objective existence of 
material systems ... which possess properties independently of measurements"} is accepted 
as an axiom.  The respective axiom implies the idea that a measurement aims to give 
information about the pre-existent state of the investigated system. We opine that the 
mentioned axiom and idea must also be adopted  in connection with the quantum systems. Our
opinion is encouraged by the following two remarks expressed with regard to QMS:

(i)\emph{"when it is said that something is \emph{"measured"} it is difficult not to
 think of the results are referring to some preexisting property of the object in question"}\cite{72}.\\
(ii)\emph{"the function of measurement in quantum mechanics is to determine the average 
value and the dispersion of some physical quantity in a given system as they 
are prior the measurement"}\cite{76}.

For discussions on the QMS it is also important to underline the fact that quantum 
observables are stochastic variables. Then, as it was pointed out in \textbf{R.46}, in the last 
analysis a true measurement of such an observable requires a statistical sampling (composed from
a large number of single experimental trials). A similar requirement is found  in the 
case of measurements regarding the measurement of  classical (non-quantum) observables
with stochastic characteristics. That is why we think that in the description of QMS 
there can be much interest in some ideas referring to the measurements of the alluded 
classical observables (like the ones discussed in \cite{77,78} ).

The above considerations plead for the idea that for the description of QMS a naturally 
reconsidered approach can be founded on the following new premises (\textbf{\emph{NP}}):
     
\textbf{\emph{\underline{NP.1}}}: \emph{The real purpose of a QMS is to give information 
about the pre-existent state of the investigated system. Therefore for a natural description 
of a QMS it is needlessly any reference to a collapse (reduction) of the respective state
into a post-measurement one.}

\textbf{\emph{\underline{NP.2}}}: \emph{The description of QMS has to assimilate some ideas 
regarding the measurements of classical stochastic observables. This is because, 
in practice, both types of measurements consist in similar statistical samplings.}

\textbf{\emph{\underline{NP.3}}}: \emph{Since QMS refer to the systems studied in QM their 
descriptions ought to use some QM concepts (e.g. wave functions and operators).} 

\textbf{\emph{\underline{NP.4}}}: \emph{The procedures of QMS include parts which do not
belong to QM. Therefore the description of QMS have to be regarded not as a part of QM-theory 
but as a distinct scientific branch.}

By taking into account the above premises \textbf{\emph{NP.1 - 4}} a new approach of QMS description 
can be developed, which will be presented below in the next sections.

\section{A suggesting classical model} \label{sec10}
An approach of QMS, reconsidered in the above mentioned senses, can be started 
with some discussions about a suggesting model which regards the measurements of 
classical stochastic observables. For such a model we refer to the case of an observable 
$\rm A$ appearing in the phenomenological theory of fluctuations \cite{65}, \cite{79}
(mentioned also in the remark \textbf{R.46} ). The values $a$ of such an observable
range in a continuous and infinite spectrum (i.e. $a \in (-\infty, \infty)$). The 
respective values are associated with the probability distribution $w(a)$. In respect 
with the values of $\rm A$ a measurement can be regarded \cite{76}, \cite{78} as an 
input $\rightarrow$ output ($in \rightarrow out$) process which imply the transformation
\begin{equation}\label{eq:60}
w_{in} \left( a \right) \to w_{out} \,\left( a \right)
\end{equation}
In the alluded regard $w_{in} \,\left( a \right)$ refers to the intrinsic values of 
$\rm A$, associated with the inner properties of the measured system, while 
$w_{out} \,\left( a \right)$ is related with the output values of $\rm A$, displayed 
on the recorder of the measuring device. Note that, without any loss of generality, 
the $in$ and $out$ spectra of $\rm A$ (associated to $w_{in} \,\left( a \right)$ and
$w_{out} \,\left( a \right)$) can be considered as coincident. By means of 
$w _\eta (a)$ ($\eta = in, out$)  the corresponding (numerical) characteristics
of $\rm A$ regarded as stochastic  variable can be introduced. In the spirit of 
usual practice of physics we refer here only to the two lowest order such characteristics. 
They are the $\eta$ - mean (expected) values $\left\langle {\rm A} \right\rangle _\eta $ and $\eta$ 
- standard deviations $\Delta _\eta {\rm A}$ defined as follows
\begin{equation}\label{eq:61}
\left\langle {\rm A} \right\rangle _\eta  = \int\limits_{ - \infty }^\infty  
{a\,w_\eta } \left( a \right)da \quad \left( {\Delta _\eta {\rm A}} \right)^2  = 
\left\langle {\left( {{\rm A} - \left\langle {\rm A} \right\rangle _\eta } \right)^2 } 
\right\rangle _\eta 
\end{equation}
The values of $\rm A$ displayed by the measuring device are evaluated (by sampling and
processing) according to the rules of mathematical statistics \cite{83,36}. 
As regards the observable $\rm A$ the mentioned evaluation attends to provide optimal estimators
for the $out$-characteristics like $\left\langle {\rm A} \right\rangle _{out} $ and 
$\Delta _{out} {\rm A}$ (or even for the $out$-distribution $w _{out}(a)$). It is a fact 
that some publications seem to admit (tacitly) the idea that the alluded evaluation 
takes into account the $in$-characteristics of $\rm A$ (like $\left\langle {\rm A} \right\rangle _{in} $, $\Delta _{in}\rm A $ 
or even $w _{in} (a)$). But one can see that such an  idea is justified only in 
the case of ideal measurements when $w _{out} (a) = w _{in} (a)$. However  in most of the 
real situations the measurements are non-ideal and $w _{out} (a) \neq w _{in} (a)$. 
Consequently we think that, for the considered measurements of $\rm A$, it is necessary to search a
theoretical description able to give a concrete and credible expression for the 
transformation \eqref{eq:60}. In the spirit of our works \cite{77}, \cite{78} the 
respective search can be materialized by taking into account the following aspects:

(i) For a characterization of the measuring device we use the transfer probability
$G (a, a')$ with the significance : (i.1) $G (a, a')da$ denote the (infinitesimal)
probability that by measurement the $in$-value $a'$ of $\rm A$ to be recorded in the 
$out$-interval $(a, a+da)$,  (i.2) $G (a, a')da'$ represents the probability
that the $out$-value $a$ to result from the $in$-values which belong to the interval 
$(a', a'+da')$.\\
(ii) The stochastic characteristics of the measuring device and of the measured 
system  respectively are completely independent.

According to the rules of composition for probabilities the transformation 
\eqref{eq:60} can be written as 
\begin{equation}\label{eq:62}
w_{out} \left( a \right) = \int\limits_{ - \infty }^\infty  {G\left( {a,\,a'}
\right)} \,w_{in} \left( {a'} \right)da'
\end{equation}
Due to the significances  mentioned above in (i.1) and (i.2) the kernel $G (a, a')$ 
satisfies the relation  
\begin{equation}\label{eq:63}
\int\limits_{ - \infty }^\infty  {G\left( {a,\,a'} \right)} \,da = \int\limits_{ 
- \infty }^\infty  {G\left( {a,\,a'} \right)} \,da' = 1
\end{equation}
One observes that in \eqref{eq:62} $G (a, a')$ describe the characteristics of the
measuring devices. Particularly $G (a, a') = \delta (a - a')$ and 
$G (a, a') \neq \delta (a - a')$ (where $\delta$ denotes the Dirac function) describe 
an ideal respectively a non-ideal device. Differently $w_{in} \left( {a'} \right)$
refers to the properties of the measured system. So  $w_{out} \left( {a} \right)$ 
incorporate information regarding both the mentioned device and system.

Now, from the general perspective  of the present paper, it is of interest to note some 
observations about the measuring uncertainties (errors). Firstly it is important to 
remark that for the discussed observable $\rm A$, the standard deviations 
$\Delta _{in} \rm A$ and $\Delta _{out} \rm A$ are not estimators of the  mentioned
uncertainties. Of course that the above remark contradicts some loyalities induced by CIUR
doctrine. Here it must be pointed out  that: 

(i) On  one hand $\Delta _{in} \rm A$ together with 
$\left\langle \rm A \right\rangle _{in}$ describe only the 
intrinsic properties of the measured system.\\
(ii) On the other hand $\Delta _{out} \rm A$ and $\left\langle \rm A \right\rangle _{out}$
incorporate composite information about the respective system and the measuring device.

Then, in terms of the above considerations, the measuring uncertainties of $\rm A$ are
described by the following error indicators (characteristics)
\begin{equation}\label{eq:64}
\varepsilon \left\{ {\left\langle {\rm A} \right\rangle } \right\} = \left| 
{\left\langle {\rm A} \right\rangle _{out}  - \left\langle {\rm A} \right\rangle _{in} } 
\right|\;,\quad \varepsilon \left\{ {\Delta \,{\rm A}} \right\} = \left| {\Delta _{out} 
{\rm A} - \Delta _{in} {\rm A}} \right|
\end{equation}
Note that because $\rm A$ is  a stochastic  variable for an acceptable evaluation of its
measuring uncertainties it is completely insufficient the single indicator
$\varepsilon\left\{ {\left\langle {\rm A} \right\rangle } \right\}$. Such an evaluation 
requires at least the couple $\varepsilon \left\{ {\left\langle {\rm A} \right\rangle } \right\}$and
$\varepsilon \left\{ {\Delta \,{\rm A}} \right\}$ or even the differences  of the higher 
order moments like 
\begin{equation}\label{eq:65}
\varepsilon \left\{ {\left\langle {\left( {\delta {\rm A}} \right)^n } \right\rangle }
\right\} = \left| {\left\langle {\left( {\delta _{out} {\rm A}} \right)^n } 
\right\rangle _{out}  - \left\langle {\left( {\delta _{in} {\rm A}} \right)^n } 
\right\rangle _{in} } \right| 
\end{equation}
where $\delta _\eta  {\rm A} = {\rm A} - \left\langle {\rm A} \right\rangle _\eta\,; 
\eta  = in,\;out\,; \; n \ge 3  $).

Add here the observation that a comprehensive characterization of the measuring 
uncertainties regarding $\rm A$ can be done \cite{77} in terms of informational (Shannon)
entropies $S(w _\eta)$ defined as
\begin{equation}\label{eq:66}
S\,\left( {w_\eta } \right) =   -  \int\limits_{ - \infty }^\infty  {w_\eta \left( a 
\right)\,ln\,w_\eta \left( a \right)da\,} , \quad \left( {\eta = in,\,out} \right)
\end{equation}
By using the relations \eqref{eq:62} and \eqref{eq:63}  together with the evident 
formula $ x - 1 \geq ln (x)$ (with $x \in (0, \infty)$) it is easy to prove \cite{77} the 
following result
\begin{equation}\label{eq:67}
\varepsilon \left\{ {S\left( w \right)} \right\} = S\left( {w_{out} } 
\right) - S\left( {w_{in} } \right) \ge 0
\end{equation}
This result shows that during the measurement the error $\varepsilon \left\{ 
{S\left( w \right)} \right\}$ of the informational 
entropy (associated with the observable $\rm A$) is a real and non-negative quantity.
In the last part of \eqref{eq:67} the signs $=$ and $>$ correspond to an ideal and a 
non-ideal measurement respectively.

\section{The suggested quantum model}\label{sec11}
Now let us develop  a reconsidered model for description of QMS. The announced development 
conforms oneself to the premises $\textbf{\emph{NP.1 - 4}}$ and assimilates some ideas suggested by the 
classical model discussed in the previous section. We restrict our model only to the 
measurements of quantum observables of orbital nature (i.e. coordinates, momenta, angles,
angular momenta and energy). The respective observables are described by the operators
$\hat A_j \left( {j = 1,\,2,\, \ldots ,n} \right)$ regarded as generalized stochastic
variables. As a measured system we consider a spinless microparticle whose state is 
described by the wave function  $\psi  = \psi \left( {\vec r}, t \right)$, taken in the
coordinate representation ($\vec r$ and $t$ stand for microparticle's position and time respectively). 
Account here the fact that, because we consider only  a non-relativistic context, the explicit mention of time
$t$ in the expression of $\psi$ is unimportant.

Now note the observation that the wave function $\psi \left( {\vec r} \right)$
incorporate information (of probabilistic nature) about the measured system. That is why 
a QMS can be regarded as a process of information transmission: from the respective system
to the recorder of the measuring device. Then, on the one hand, the input ($in$) information 
described by $\psi _{in} \left( {\vec r} \right)$ refers to the intrinsic (own)properties
of the measured system (regarded as information source). The expression of 
$\psi _{in} \left( {\vec r} \right)$ is deducible within the framework of usual QM (e.g. by
solving the adequate Schrodinger equation). On the other hand the output ($out$) 
information, described by the wave function  $\psi _{out} \left( {\vec r} \right)$, refers 
to the data obtained on the device recorder (regarded as information receiver). So the 
measuring device plays the role of the transmission  channel  for the alluded
information. Accordingly the measurement appears as a  processing information operation.
By regarding the things as above the description of the QMS must be associated with the
transformation
\begin{equation}\label{eq:68}
\psi _{in} \left( {\vec r} \right)  \rightarrow  \psi _{out} \left( {\vec r} \right)  
\end{equation}
As in the classical model (see the previous section) , without any loss of generality,
here we suppose that the quantum observables have identical  spectra of values  in both
$in$- and $out$- situations. In terms of QM the mentioned supposition means that 
the operators $\hat A _j$ have the same mathematical expressions in both $in$- and $out$- 
readings.
The respective expressions are the known ones from QM.

In the framework delimited by the above notifications the description of QMS requires   
putting   the transformation \eqref{eq:68} in concrete forms and then using some of the 
known rules of QM.  In our opinion the mentioned requirement must be formulated in terms of quantum
probabilities carriers. Such carriers are the probabilistic densities $\rho _\eta$ and 
currents $\vec J _\eta$ defined by
\begin{equation}\label{eq:69}
\rho _\eta  = \left| {\psi _\eta } \right|^2 \,,\quad \quad \vec J_\eta  = 
\frac{\hbar }{m}\,\left| {\psi _\eta } \right|^2 \, \cdot \nabla \phi _\eta 
\end{equation}
Here $\left| {\psi _\eta } \right|$ and $\phi _\eta$ represents the modulus and the 
argument of $\psi _\eta$ respectively (i.e.$\psi _\eta = \left| {\psi _\eta } 
\right| exp(i \phi _\eta)$) and $m$ denotes the mass of microparticle.

The alluded formulation is connected with the observations \cite{84} that the couple 
$\rho$ -  $\vec J$ \emph{"encodes the probability distributions of quantum mechanics"} and
it \emph{"is in principle measurable by virtue of its effects on other systems"}. To be
added here the possibility \cite{85} of taking in QM as primary entity the couple
$\rho _{in} - \vec J _{in}$  but not the wave function $\psi _{in}$ (i,e. to start the
QM considerations with the continuity equation for the mentioned couple  
and subsequently to derive the Schrodinger equation for $\psi _{in}$).

According to  the above observations the transformations \eqref{eq:68} have to be 
formulated in terms of  $\rho _\eta$ and $\vec J _\eta$. But   $\rho _\eta$ and  
$\vec J _\eta$ refer to the position and the motion kinds of probability respectively. 
Experimentally the two kinds can be regarded as measurable by distinct devices and procedures. Consequently the
mentioned formulation has to combine the following two distinct transformations
\begin{equation}\label{eq:70}
\rho _{in}  \to \rho _{out} \,,\quad \quad \vec J_{in}  \to \vec J_{out} 
\end{equation}
The considerations about the classical relation \eqref{eq:62} suggest that,by completely 
similar arguments,  the transformations \eqref{eq:70} admit the following 
transformations 
\begin{equation}\label{eq:71}
\rho _{out} \left( {\vec r} \right) = \iiint {\Gamma \left( {\vec r,\vec r'}
\right)\,\rho _{in} \left( {\mathord{\buildrel{\lower3pt\hbox{$\scriptscriptstyle
\rightharpoonup$}} \over r} '} \right)} \,d^3 \vec r'
\end{equation}
\begin{equation}\label{eq:72}
J_{out;\;\alpha }  = \sum\limits_{\beta  = 1}^3 {\iiint {\Lambda _{\alpha 
\beta } } } \left( {\vec r,\vec r'} \right)\,J_{in;\;\beta } \left( {\vec r'} 
\right)\,d^3 \vec r'
\end{equation}
In \eqref{eq:72} $J _{\eta; \alpha}$ with $\eta = in, out$ and $\alpha = 1,2, 3 = x,y,z$ 
denote Cartesian components of $\vec J _\eta$.

Note the fact that the kernels $\Gamma$ and $\Lambda _{\alpha \beta}$ 
from \eqref{eq:71} and \eqref{eq:72} have significance of 
transfer probabilities, completely analogous 
with the meaning  of the classical kernel $G (a, a')$ from \eqref{eq:62}. This fact 
entails the following relations
\begin{equation}\label{eq:73}
\iiint {\Gamma \left( {\vec r,\vec r'} \right)\,} d^3 \vec r = \iiint 
{\Gamma \left( {\vec r,\vec r'} \right)} \,d^3 \vec r'= 1
\end{equation}
\begin{equation}\label{eq:74}
\,\sum\limits_{\alpha  = 1}^3 {\iiint {\Lambda _{\alpha \beta } } } \left( 
{\vec r,\vec r'} \right)\,d^3 \vec r = \sum\limits_{\beta  = 1}^3 {\iiint 
{\Lambda _{\alpha \beta } } } \left( {\vec r,\vec r'} \right)\,d^3 \vec r'= 1
\end{equation}
The kernels $\Gamma$ and $\Lambda _{\alpha \beta}$  describe the transformations induced 
by QMS in the data (information) about the measured system (microparticle). Therefore they 
incorporate some extra-QM elements regarding the characteristics of measuring devices and
procedures. The respective elements do not belong to the usual QM framework which refers
to the intrinsic (own) characteristics of the measured system.

The above considerations facilitate an evaluation of the effects induced by QMS on the 
probabilistic estimators of  here considered orbital observables $A _j$. Such 
observables are described by the operators $\hat A_j $ whose expressions depend on 
$\vec r$ and $\nabla$. According to the previous discussions the mentioned operators 
are supposed to remain invariant under the transformations which describe QMS. So one can
say that in the situations associated with the wave functions $\psi _\eta$ 
($\eta = in, out$) the mentioned observables are described by the following probabilistic 
estimators/characteristics  (of  first order): mean values $\left\langle {A_j } \right\rangle _\eta $, 
correlations $C_\eta \left( {A_j ,\,A_k } \right)$ and standard deviations $\Delta _\eta A_j$. 
With the usual notation
$\left( {f,g} \right)\, = \,\int {f^* } g\,d^3 \vec r$ for the scalar product of
functions $f$ and $g$, the mentioned estimators are defined by the relations
\begin{equation}\label{eq:75}
\begin{split}
&\left\langle {A_j } \right\rangle _\eta  = \left( {\psi _\eta ,\,\hat A_j 
\psi _\eta } 
\right)\,,\qquad \qquad \delta _\eta \hat A_j = \hat A_j  - 
\left\langle {A_j } \right\rangle _\eta  \\ 
&C_\eta \left( {A_j ,\,A_k } \right) = \left( {\delta _\eta \hat A_j \,\psi _\eta 
,\,\delta _\eta \hat A_k \,\psi _\eta } \right)\,,\quad \Delta _\eta A_j = \sqrt 
{C_\eta \left( {A_j ,\,A_j } \right)} 
\end{split}
\end{equation}
Add here the fact that the $in$ version of the estimators \eqref{eq:75} are calculated 
by means of the wave function $\psi _{in}$, known from the considerations about the  
inner properties of the investigated system (e.g. by solving the corresponding Schr\"odinger 
equation). On the other hand the $out$ version of the respective estimators can be 
evaluated by  usage of the probability density and current $\rho _{out}$ and 
$\vec J _{out}$. So if $\hat A _j$ does not depend on $\nabla$ (i.e. $\hat A _j  = A _j(\vec r)$) 
in evaluating the scalar products from \eqref{eq:75} one can use the evident equality
$\psi _{out} \hat A_j \,\psi _{out}  = \hat A_j \,\rho _{out} $. When 
$\hat A _j$ depends on $\nabla$ (i.e. $\hat A _j = A _j (\nabla)$) in the same products 
can be appealed the substitution
\begin{equation}\label{eq:76}
\psi _{out}^* \,\nabla \psi _{out}  = \frac{1}{2}\nabla \rho _{out}  + 
\frac{{im}}{\hbar }\,\vec J_{out} 
\end{equation}
\begin{equation}\label{eq:77}
\psi _{out}^* \,\nabla ^2 \psi _{out}  = \rho _{out}^{\frac{1}{2}} \,
\nabla ^2 \,\rho _{out}^{\frac{1}{2}}  + 
\frac{{im}}{\hbar }\,\nabla \vec J_{out}  - \frac{{m^2 }}
{{\hbar ^2 }}\,\frac{{\vec J_{out} ^2 }}{{\rho _{out} }}
\end{equation}
The mentioned usage seems to allow the avoidance of the implications regarding \cite{84}
\emph{" a possible nonuniqueness of current"}(i.e. of the couple 
$\rho _\eta - \vec J _\eta$).

Within the above presented model of QMS the instrumental uncertainty (errors) associated 
with the measurements   of observables $A _j$ can be evaluated through the following uncertainty
indicators
\begin{equation}\label{eq:78}
\begin{split}
\varepsilon \left\{ {\left\langle {A_j } \right\rangle } \right\} &= \left| 
{\left\langle {A_j } \right\rangle _{out}  - \left\langle {A_j } 
\right\rangle _{in} } \right| \\ 
\varepsilon \left\{ {C\left( {A_j ,A_k } \right)} \right\} &= \left| {C_{out} 
\left( {A_j, A_k } \right) - C_{in} \left( {A_j ,A_k } \right)} \right| \\ 
\varepsilon \left\{ {\Delta \,A_j } \right\} &= \left| {\Delta _{out} 
\,A_j  - \Delta _{in} \,A_j } \right|
\end{split}
\end{equation}
These quantum indicators are completely similar with the classical ones \eqref{eq:64}.

Note that here can be used a similarity with the classical situation discussed in the 
previous section. So the quantum measuring uncertainties can be evaluated not only by 
the quantities \eqref{eq:78} but also through the changes in informational entropies. 
In quantum cases the respective entropies can be defined as
\begin{equation}\label{eq:79}
S\left( {\rho _\eta  } \right)= - \iiint {\rho _\eta  } \left( {\vec r} \right) 
\cdot \ln \rho _\eta  \left( {\vec r} \right)d^3 \vec r
\end{equation}
\begin{equation}\label{eq:80}
S\left( {\vec j_\eta  } \right) =  - \iiint \upsilon^{-1}|\vec j_\eta  \left( {\vec r} 
\right)|  \cdot \ln \left| \upsilon^{-1}\vec j_\eta  \left( {\vec r} \right) \right|d^3
\vec r
\end{equation}
In the last of these relations the factor $\upsilon^{-1}$ was introduced together 
with $\vec J _\eta$ because of dimensional considerations ($\upsilon$ has the dimension
of velpcity ).

Then in an informational view the measuring uncertainties are described by the couple
of the   indicators $\varepsilon \left\{ {S\left( \rho  \right)} \right\}$ and 
$\varepsilon \left\{ {S\left( {\vec J} \right)} \right\}$defined by the following
relations :
\begin{equation}\label{eq:81}
\varepsilon \left\{ {S\left( \rho  \right)} \right\} = \left| {S\left( 
{\rho _{out} } \right) - S\left( {\rho _{in} } \right)} \right| \ge 0
\end{equation}
\begin{equation}\label{eq:82}
\varepsilon \left \{ {S \left( {\vec J} \right)} \right\}= \left| 
{S\left( {\vec J_{out} } \right)\; - \;S\left( {\vec J_{in} } \right)} \right|
\ge 0
\end{equation}
The  relation \eqref{eq:81} can be proved similarly with \eqref{eq:67}. Probably that 
proof of relation \eqref{eq:82}   requires more elaborate mathematical reasonings 
(it was written here on intuitive considerations). In both of the respective relations 
the sign $=$ refers to the ideal measurements  while the cases with $>$ regard the 
non-ideal measurements. \\
The here discussed model regarding the description of QMS is exemplified in Annex A.

Now is the place to note that the $out$-version of the estimators \eqref{eq:75}, 
\eqref{eq:79} and \eqref{eq:80} as well as the uncertainty indicators \eqref{eq:78}, 
\eqref{eq:81} and \eqref{eq:82} have a theoretical significance. In practice the 
verisimilitude of such estimators and indicators must be tested by
comparing them with their experimental correspondents (obtained by sampling and processing
of the data collected from the recorder of the measuring device). If the test is 
confirmative both theoretical descriptions, of QM intrinsic properties of system and 
of QMS, can be considered as adequate. But if the test gives an invalidation of the 
results, at least one of the mentioned descriptions must be regarded as inadequate.

In the end of this section we wish to add the following two observations:

(i) The here proposed  description of QMS does not imply some interconnection of principle 
between the measuring uncertainties of two distinct observables. This means that from 
the perspective of the respective description there are no reasons to discuss about a 
measuring compatibility or incompatibility of two observables.\\
(ii) The above considerations from the present section
refer to the QMS of orbital observables. Similar considerations can be also  done  in the
case of QMS regarding the spin observables. In such  a case besides the probabilities
of spin-states (well known in QM publications) it is important to take into account
the spin current density (e.g. in the version proposed recently \cite{86}).

\section{Some conclusions}\label{sec12}
We starred the present paper from the observation that in fact CIUR is troubled by a 
number of still unsolved shortcomings. Then, for a primary goal of our text, we strove
to investigate in detail the main aspects as well as the  authenticity of the
respective shortcomings. So we firstly analysed  the renowned  deficiencies regarding the
pairs of canonically conjugate  observables $L _z - \phi$, $N - \phi$ and $E - t$.
Additionally we also  discussed a whole class of other CIUR shortcomings usually 
underestimated (or even neglected) in publications.

The mentioned investigations, performed in sections 2 - 6, reveal the following 
aspects :

(i) A group of the CIUR's shortcomings appear from the application of the usual 
(Robertson -Schrodinger) version of UR in situations where, mathematically, it is 
incorrect ;\\
(ii) The rest of the shortcomings result from unnatural linkages with things of other 
nature (e.g. with the thought experimental relations or with the presence/absence of                $\hbar$ in some formulas);\\
(iii) Moreover one finds that, if the mentioned applications and linkages are 
handled correctly, the alluded shortcomings prove themselves as being veridic and 
unavoidable facts. The ensemble of the respective facts invalidate all the 
basic ideas of CIUR.

In consensus with the above noted findings, in Section 7, we promoted the opinion that
CIUR must be abandoned as an incorrect and useless (or even misleading) doctrine. 
Conjointly with the respective opinion we think that the primitive  UR (the so called 
 Heisenberg's relations) must be regarded as:

(i) fluctuation formulas - in their theoretical (Robertson-Schrodinger) version,\\
(ii) fictitious things, without any physical significance - in their thought-experimental
version.

Because CIUR ideas imply suppositions regarding QMS the above announced abandonment
requires a re-examination ( at least in part) of the QMS problems. To such a requirement we tried
to answer in Sections 8 - 11. So, by a detailed investigation, we have shown that the
CIUR-connected approaches of QMS are grounded on dubitable (or even incorrect) premises.
That is why we declare ourselves for reconsidered approaches of QMS, based on new
(and more natural) premises. Such an approach is argued and developed in Sections 9, 10 and
11 respectively, and it is exemplified in Annex A.  

\section*{Acknowledgments}\ 
\begin{itemize}
\item  I Wish to express my deep gratitude to those authors, publishing companies 
and libraries which, during the years, helped me with  copies of some publications
connected with the problems approached here.
\item The investigations in the field of the present paper benefited partially 
of facilities from the grants supported by the Roumanian Ministry of Education and
Research.
\end{itemize}

 \section*{Annex A: An exemplification}
For a simple exemplification of the model presented in Section 11 let us refer to a
microparticle in a one-dimensional motion along the $x$-axis. We take 
$\psi _{in} \left( x \right) = \left| {\psi _{in} \left( x \right)} \right| 
\cdot \exp \left\{ {i\phi _{in} \left( x \right)} \right\}$ 
with
\begin{equation}\label{eq:83}
\left| {\psi _{in} \left( x \right)} \right| = \left( {\sigma \sqrt {2\pi } }
\right)^{ - \frac{1}{2}}  \cdot \exp \left\{ { -  \frac{{\left( {x - x_0 } 
\right)^2 }}{{4\sigma ^2 }}} \right\}\,,\quad \phi _{in} \left( x \right) =kx
\end{equation}
Correspondingly we have
\begin{equation}\label{eq:84}
\rho _{in} \left( x \right) = \left| {\psi _{in} \left( x \right)} \right|^2 
\,,\quad \quad J_{in} \left( x \right) = \frac{{\hbar k}}{m}\,\left| {\psi _{in} 
\left( x \right)} \right|^2 
\end{equation}
So the intrinsic properties of the microparticle are described by the parameters
$x _0$, $\sigma$  and $k$.

If the errors induced by QMS are small the kernels $\Gamma$ and $\Lambda$ in 
\eqref{eq:71}-\eqref{eq:72} can be considered of Gaussian forms like
\begin{equation}\label{eq:85}
\Gamma \left( {x,x'} \right) = \left( {\gamma \sqrt {2\pi } } \right)^{ - 1} 
\cdot \exp \left\{ { - \frac{{\left( {x - x'} \right)^2 }}{{2\gamma ^2 }}} \right\}
\end{equation}
\begin{equation}\label{eq:86}
\Lambda \left( {x,x'} \right) = \left( {\lambda \sqrt {2\pi } } \right)^{ - 1} 
\cdot \exp \left\{ { - \frac{{\left( {x - x'} \right)^2 }}{{2\lambda ^2 }}} \right\}
\end{equation}
where $\gamma$ and $\lambda$ describe the characteristics of the measuring devices. Then
for $\rho _{out}$ and $J _{out}$one finds
\begin{equation}\label{eq:87}
\rho _{out} \left( x \right) = \left[ {2\pi \left( {\sigma ^2  + \gamma ^2 }
\right)} \right]^{ - \frac{1}{2}}  \cdot \exp \left\{ { - \frac{{\left( {x - x'}
\right)^2 }}{{2\left( {\sigma ^2  + \gamma ^2 } \right)}}} \right\}
\end{equation}
\begin{equation}\label{eq:88}
J_{out} \left( x \right) = \hbar k \left[ {2\pi m^2 \left( {\sigma ^2  + 
\lambda ^2 } \right)} \right]^{ - \frac{1}{2}}  \cdot \exp \left\{ { - \frac{{\left( {x - x'} 
\right)^2 }}{{2\left( {\sigma ^2  + \lambda ^2 } \right)}}} \right\}
\end{equation}
It can been seen that in the case when both $\gamma \rightarrow 0$ and
$\lambda \rightarrow 0$ the kernels $\Gamma (x,x')$ and $\Lambda (x,x')$ degenerate 
into the Dirac's function $\delta (x-x')$. Then $\rho _{out} \rightarrow \rho _{in}$ and
$J _{out} \rightarrow J _{in}$.  Such a case corresponds to an ideal measurement. 
Alternatively the cases when $\gamma \neq 0$ and/or $\lambda \neq 0$ are associated with
non-ideal measurements.

As observables of interest we take coordinate $x$ and momentum $p$ described by the operators
$\hat x = x \cdot$ and $\hat p =  - i\hbar \frac{\partial }{{\partial x}}$.
Then according to the scheme presented in Section \ref{sec11} one obtains
\begin{equation}\label{eq:89}
\left\langle x \right\rangle _{in} = \left\langle x \right\rangle _{out}
= x_0 \,,\quad \left\langle p \right\rangle _{in}  = \left\langle p 
\right\rangle _{out}  = \hbar k
\end{equation}
\begin{equation}\label{eq:90}
C_{in} \left( {x,p} \right) = C_{out} \left( {x,p} \right) = 
\frac{{i\hbar }}{2}
\end{equation}
\begin{equation}\label{eq:91}
\Delta _{in} x = \sigma \,,\quad \quad \Delta _{out} x = \sqrt {\sigma ^2  + \gamma ^2 } 
\end{equation}
\begin{equation}\label{eq:92}
\Delta _{in} p = \frac{\hbar }{{2\sigma }}
\end{equation}
\begin{equation}\label{eq:93}
\Delta _{out} p = \hbar \sqrt {\frac{{k^2 \left( {\sigma ^2  + \gamma ^2 } 
\right)}}{{\sqrt {\left( {\sigma ^2  + \lambda ^2 } \right)\left( {\sigma ^2  + 
2\gamma ^2  - \lambda ^2 } \right)} }} - k^2  + \frac{1}{{4\left( {\sigma ^2  + 
\gamma ^2 } \right)}}} 
\end{equation}
Subsequently, according to \eqref{eq:78}, the measuring errors regarding $x$ and $p$
are characterized by the uncertainty indicators
\begin{equation}\label{eq:94}
\varepsilon \left\{ {\left\langle x \right\rangle } \right\} = 0\,,\quad 
\varepsilon \left\{ {\left\langle p \right\rangle } \right\} = 0\,,\quad \varepsilon \left\{ 
{C\left( {x,p} \right)} \right\} = 0
\end{equation}
\begin{equation}\label{eq:95}
\varepsilon \left\{ {\Delta \,x} \right\} = \sqrt {\sigma ^2  + \gamma ^2 } 
- \sigma 
\end{equation}
\begin{equation}\label{eq:96}
\varepsilon \left\{ {\Delta \,p} \right\} = \left| {\Delta _{out} \,p - \Delta _{in}
\,p} \right|
\end{equation}
(Read the last formula by appealing to \eqref{eq:92} and \eqref{eq:93}).

The relations \eqref{eq:94}-\eqref{eq:96} show that in the considered model the
characteristics ${\left\langle x \right\rangle }$, ${\left\langle p \right\rangle }$ and
$C (x,p)$ of stochastic observables $x$ and $p$ are not troubled by measuring errors.
But in the same model the characteristics $\Delta x$ and $\Delta p$ are disturbed by
non-null such errors.

For the measuring uncertainties \eqref{eq:81} and \eqref{eq:82} regarding the informational 
entropies one finds
\begin{equation}\label{eq:97}
\varepsilon \left\{ {S\left( \rho  \right)} \right\} = S\left( {\rho _{out} } 
\right) - S\left( {\rho _{in} } \right) = \frac{1}{2}\ln \left( 
{1 + \frac{{\gamma ^2 }}{{\sigma ^2 }}} \right)
\end{equation}
\begin{equation}\label{eq:98}
\varepsilon \left\{ {S\left( J \right)} \right\} = S\left( {J_{out} } 
\right) - S\left( {J_{in} } \right) = \frac{1}{2}\ln \left( 
{1 + \frac{{\lambda ^2 }}{{\sigma ^2 }}} \right)
\end{equation}
In the last of these relation for $\upsilon$ introduced in \eqref{eq:80} we take
$\upsilon = (m/\hbar k)$.

For an evaluation of the interdependence between the uncertainty indicators of $x$ and
$p$ from \eqref{eq:94}-\eqref{eq:96} one obtains
\begin{equation}\label{eq:99}
\varepsilon \left\{ {\left\langle x \right\rangle } \right\} \cdot \varepsilon \left\{ 
{\left\langle p \right\rangle } \right\} = 0
\end{equation}
\begin{equation}\label{eq:100}
\varepsilon \left\{ {\Delta x} \right\} \cdot \varepsilon \left\{ {\Delta p} \right\} 
\ge \hbar \mu 
\end{equation}
Here $\mu$ is a real, non-negative and dimensionless quantity which can be evaluated
by means of the relations \eqref{eq:94}-\eqref{eq:96}.

If in \eqref{eq:83} we restrict to the values $x _0 = 0$, $k = 0$ and
$\sigma = \sqrt {\frac{\hbar }{{2m\omega }}} $ our system is just a linear 
oscillator in its ground state ($m$ = mass and $\omega$ = angular frequency). As observable
of interest we consider the energy  described by the Hamiltonian
\begin{equation}\label{eq:101}  
\hat {\rm H} =  - \frac{{\hbar ^2 }}{{2m}}\,\frac{{d^2 }}{{d{\kern 1pt} x^2 
}} + \frac{{m\,\omega ^2 }}{2}\,x^2 
\end{equation}
Then for the respective observable one finds
\begin{equation}\label{eq:102}
\left\langle {\rm H} \right\rangle _{in}  = \frac{{\hbar \,\omega }}{2}\,,
\quad \quad \Delta _{in} {\rm H}= 0
\end{equation}
\begin{equation}\label{eq:103}
\left\langle {\rm H} \right\rangle _{out}  = \frac{{\omega \left[ {\hbar ^2  + 
\left( {\hbar  + 2m\,\omega \,\gamma ^2 } \right)^2 } \right]}}{{4\left( {\hbar  
+ 2m\,\omega \,\gamma ^2 } \right)}}
\end{equation}
\begin{equation}\label{eq:104}
\Delta _{out} \,{\rm H} = \frac{{\sqrt 2 \,m\,\omega ^2 \,\gamma ^2 \left( 
{\hbar  + m\,\omega \,\gamma ^2 } \right)}}{{\left( {\hbar  + 2m\,\omega \,\gamma ^2 } 
\right)}}
\end{equation}
The corresponding instrumental errors are described by the uncertainty indicators
\begin{equation}\label{eq:105}
\varepsilon \left\{ {\left\langle {\rm H} \right\rangle } \right\} = \left| 
{\left\langle 
{\rm H} \right\rangle _{out}  - \left\langle {\rm H} \right\rangle _{in} }
\right| \ne 0
\end{equation}
\begin{equation}\label{eq:106}
\varepsilon \left\{ {\Delta {\rm H}} \right\} = \left| {\Delta _{out} \,
{\rm H} - \Delta _{in} {\rm H}} \right| \ne 0
\end{equation}

\section*{List of abreviations}
\textbf{\emph{BI}}= basic ideas \\   
CIUR = conventional interpretation of uncertainty relations\\
\textbf{\emph{CP}}	= conventional premises \\
EXR = extended rotations \\
$in$ = input\\
\textbf{\emph{NP}}	= new premises \\
$out$ = output\\
QM = quantum mechanics\\
QMS = quantum measurements \\
QTP = quantum torsion pendulum\\
\textbf{R} = remark\\
SRC = sharp circular rotations\\ 
\emph{srte }= super-resolution-thought-experimental\\
\emph{te} = thought-experimental\\
UR = uncertainty relation(s)

\end{document}